\documentclass[prd,twocolumn,reprint,preprintnumbers,nofootinbib,superscriptaddress]{revtex4-2}
\input{header.tex}

\begin{document}

\reportnum{-2}{CERN-TH-2025-048}

\title{Di-decay signature of new physics particles at intensity frontier experiments}

\author{Giovani Dalla Valle Garcia}%
 \email{giovani.garcia@student.kit.edu}
\affiliation{Institut für Astroteilchenphysik, Karlsruhe Institute of Technology, Karlsruhe, Germany
}%
\author{Maksym Ovchynnikov}%
 \email{maksym.ovchynnikov@cern.ch}
\affiliation{Theoretical Physics Department, CERN, 1211 Geneva 23, Switzerland}%

\date{\today}

\begin{abstract}
We present a discovery-era strategy for Intensity Frontier experiments based on ``di-decay'' events, in which feebly interacting particles (FIPs) are produced in pairs and then both decay inside the detector. This signature may exist in broad classes of models with FIPs decaying in the same way as minimal ``portal'' particles. The two displaced vertices may allow reconstruction of the pair invariant mass and hence reveal the production channel, information normally inaccessible at these facilities. This, in turn, lets one discriminate between FIP models that share identical decay signatures. As a case study, we consider the model of Higgs-like scalars and show that SHiP, Belle II, and the Downstream@LHCb can each cover wide domains of viable parameter space using di-decays.
\end{abstract}

\maketitle

\section{Introduction}
\label{sec:intro}

Unresolved problems beyond the Standard Model (SM), such as dark matter, neutrino oscillations, and the baryon asymmetry of the Universe, may point to the existence of new particles. Particularly well-motivated candidates are relatively light states ($M \ll \Lambda_{\text{EW}}$) that interact only very feebly with the SM. Such particles are commonly referred to as Feebly-Interacting Particles (FIPs)~\cite{Alekhin:2015byh,Beacham:2019nyx,Antel:2023hkf,Asaka:2005pn,Bauer:2017ris,Aloni:2018vki,Bauer:2020jbp,Bauer:2021mvw,DallaValleGarcia:2023xhh,Ovchynnikov:2025gpx,Ilten:2018crw,Kyselov:2024dmi,Boiarska:2019jym,Ferber:2023iso,Bondarenko:2018ptm,Tsai:2019buq}. 

A broad class of experiments has incorporated searches for GeV-scale FIPs as a central component of their physics programs. Operating with high-intensity beams or collisions that can copiously produce feebly coupled particles, these facilities are collectively known as Intensity Frontier (IF) experiments~\cite{Beacham:2019nyx,Antel:2023hkf}. Prominent examples include Belle II~\cite{Belle-II:2018jsg}, NA62~\cite{NA62:2023qyn}, FASER~\cite{FASER:2019aik}, SND@LHC~\cite{SNDLHC:2022ihg}, SHiP~\cite{Aberle:2839677}, MAPP~\cite{MoEDAL-MAPP:2022kyr}, DarkQuest~\cite{Apyan:2022tsd}, DUNE~\cite{DUNE:2020lwj}, Downstream@LHCb~\cite{Gorkavenko:2023nbk,Kholoimov:2025cqe}, and many others~\cite{Bauer:2019vqk,MATHUSLA:2022sze,Aielli:2019ivi,Niedziela:2020kgq,Feng:2022inv,Hacisahinoglu:2025xrs}. These experiments probe long-lived FIPs in regions of parameter space complementary to prompt searches at the main LHC detectors, extending sensitivity toward small couplings and long lifetimes.

The maximal, signature-agnostic region in the plane of FIP parameters accessible to IF experiments is commonly referred to as their exclusion potential. In the absence of a signal, the corresponding parameter space can be ruled out. For unstable FIPs, the canonical exclusion signature is the decay of a single long-lived particle within a displaced decay volume; throughout this work, we refer to this signature as a \textit{mono-decay}.

However, the ultimate goal of IF experiments is not merely exclusion but \emph{discovery} -- namely, determining the properties of a FIP and identifying its underlying theoretical origin if a signal is observed. In favorable cases, mono-decays may allow partial characterization through reconstruction of the invariant mass, decay modes, and related observables, which can then be compared to theoretical predictions for different FIP classes~\cite{Tastet:2021vwp,Mikulenko:2023olf,Mikulenko:2023iqq,Chathirathas:2024mep,Mikulenko:2024hex,Cao:2024rzb}.\footnote{At this point, we should mention theoretical uncertainties in the decay description of GeV-mass FIPs: they can be significant, reaching orders of magnitude~\cite{Monin:2018lee,Blackstone:2024ouf,Ovchynnikov:2025gpx}, leading to ambiguities in matching data with a particular FIP model.} Nevertheless, mono-decays are often insufficient for robust model differentiation. Distinct FIPs may share identical decay phenomenology, rendering discrimination based solely on decay channels impossible.

A more powerful strategy would combine decay information with knowledge of the production mechanism. In practice, however, the production vertex at IF experiments is typically inaccessible: it occurs inside thick targets or in environments overwhelmed by backgrounds, making direct reconstruction extremely challenging or impossible.

In this work, we propose and analyze, for the first time, a complementary strategy to overcome this fundamental limitation: the observation of events featuring the simultaneous decays of two FIPs, which we refer to as \emph{di-decays}. Such events arise naturally in models that mimic the mono-decay signatures of the simplest ``portal'' benchmark models, which are widely considered by the community. We demonstrate that di-decays provide qualitatively new information about the production mechanism and thus open a pathway toward genuine FIP model differentiation.

In this work, we focus on the minimal Higgs-like scalar benchmark, which provides a clean and renormalizable setting in which to assess the phenomenological relevance and experimental reach of the di-decay signature. We investigate the discovery prospects of di-decays across a broad range of masses and couplings, and show that several IF experiments can access sizable regions of parameter space through this signature.\footnote{For the main LHC detectors, the di-decay signature has been studied in~\cite{CMS:2012qms,Curtin:2013fra,CidVidal:2019urm,ATLAS:2022izj,ATLAS:2023ian,CMS:2024jyb,CMS:2024vjn,CMS:2024zfv}. For Belle II, the di-decay signature was also discussed in Ref.~\cite{Acevedo:2021wiq}, albeit in the context of very specific models involving multiple FIPs and without revealing the discovery power.} In particular, we find that for Belle~II, the di-decay channel may even be competitive with mono-decays, owing to its large dataset and excellent coverage of the interaction point.

A second, qualitatively different benchmark based on dark QCD coupled via dark photons -- featuring pair-produced dark mesons, a distinct production mechanism, and a different invariant-mass structure -- has been studied in detail by one of the authors using the framework developed here in Ref.~\cite{Bernreuther:2025xqk}. Taken together, these works show that the di-decay strategy is not restricted to the scalar portal, but extends naturally to richer hidden-sector dynamics. A systematic program of FIP model differentiation will be pursued in future work, including further dedicated studies of representative di-decay scenarios~\cite{DallaValleGarcia:2025ta}.

Technical details required to reproduce our results are provided in Appendices.

\section{Di-decays}

In this section, we explain why di-decays provide qualitatively new information at IF experiments. We first clarify the intrinsic limitations of mono-decays for FIP identification. We then show how di-decays give direct access to the production mechanism and may lift otherwise unavoidable model degeneracies. Finally, we discuss under which parametric conditions di-decays are experimentally competitive, setting the stage for the quantitative analysis that follows.

\subsection{Di-decays as a probe of production mechanisms}

The models most commonly targeted at IF experiments are the so-called ``portal'' scenarios, in which unstable FIPs couple linearly to SM operators~\cite{Alekhin:2015byh,Beacham:2019nyx}. Examples include heavy neutral leptons (HNLs), axion-like particles (ALPs), Higgs-like scalars, dark photons (DPs), and related states. These constructions are typically minimal extensions of the SM, introducing only a single new particle.

However, mono-decays -- i.e., observing a single displaced decay vertex -- generally probe only the decay properties of the FIP. They provide access to quantities such as invariant mass, decay modes, and lifetime, but not directly to the production dynamics. This limitation becomes critical because more elaborate, yet equally motivated, extensions of the SM can reproduce exactly the same decay phenomenology. 

In particular, additional operators may modify the production of FIPs without affecting their decay channels. Since at IF experiments the production vertex is typically inaccessible -- it occurs inside thick targets or in high-background environments -- the production information is effectively lost. As a consequence, even in the case of discovery, mono-decays alone may not allow one to distinguish between qualitatively different theoretical scenarios.

This degeneracy is phenomenologically important: operators that affect production while leaving decays unchanged may fundamentally alter the nature of the FIP and its possible role in addressing problems beyond the SM.

A broad class of motivated models predicts pair production of FIPs. Pair production appears, e.g., in the following scenarios (see \cref{fig:representative_models}):
\begin{enumerate}
    \item Quadratic coupling to various SM states. A famous example is the $hXX$ operator, where $h$ is the Higgs boson, and $X$ is ALPs, Higgs-like scalars, or dark photons (DPs)~\cite{Boiarska:2019vid,Beacham:2019nyx,Antel:2023hkf,ATLAS:2022izj,CMS:2024jyb,Curtin:2013fra,Curtin:2023skh,Graesser:2007yj,Graesser:2007pc,Caputo:2017pit,Butterworth:2019iff,Barducci:2020icf,Bauer:2022rwf}.
    \item Quadratic coupling to a short-lived new physics resonance $YXX$; examples are Heavy Neutral Leptons (HNLs) interacting with the dark $U(1)'$ boson~\cite{Deppisch:2019kvs,Abdullahi:2022jlv} and different scenarios with dark matter~\cite{Berlin:2018jbm,DallaValleGarcia:2023xhh,DallaValleGarcia:2025ta}, where, e.g., Higgs-like scalars may be produced together with an unstable heavy fermion.
    \item Dark QCD~\cite{Schwaller:2015gea,Kribs:2016cew,Bernreuther:2019pfb,Albouy:2022cin,Cheng:2021kjg,Cheng:2024hvq}, where dark mesons $\pi_{D}$ and $\rho_{D}$ are typically produced with high multiplicity per event, and may mimic decays of ALPs or dark photons.
\end{enumerate}
Importantly, some of these models with di-production share the same decay phenomenology, so there is an additional degeneracy between them.\footnote{Note that such models may be involved in scenarios with dark matter, inflation, or represent the rich structure of dark sectors~\cite{Bezrukov:2009yw,Krnjaic:2015mbs,Bernreuther:2019pfb}. In this study, we are agnostic about their relation to BSM problems, though, concentrating on the experimental reach.}

\newcommand{\simplelabelshift}{24pt} 
\newcommand{\darklabelshift}{8pt}   

\begin{figure*}[t!]
\centering

\tikzset{
  darkblobpi/.style={
    draw,
    circle,
    fill=gray!25,
    inner sep=0pt,
    minimum size=8mm,
    font=\scriptsize
  },
  darkblobrho/.style={
    draw,
    circle,
    fill=gray!25,
    inner sep=0pt,
    minimum size=11mm,
    font=\scriptsize
  },
  gluonline/.style={
    decorate,
    decoration={coil, aspect=0, segment length=3.8pt, amplitude=1.45pt}
  }
}

\begin{minipage}[t][4.00cm][c]{0.30\textwidth}
\centering
\begin{tikzpicture}[baseline=(current bounding box.center), font=\Large, line width=0.95pt]
  \coordinate (h) at (-1.55,0);
  \coordinate (V)  at (0,0);
  \coordinate (X1) at (1.25, 0.95);
  \coordinate (X2) at (1.25,-0.95);

  \fill (V) circle (2.0pt);

  \draw[dashed] (h) -- (V);
  \draw (V) -- (X1);
  \draw (V) -- (X2);

  \node[left=2pt] at (h) {$h$};
  \node[right=2pt] at (X1) {$X$};
  \node[right=2pt] at (X2) {$X$};
\end{tikzpicture}

\vspace*{\simplelabelshift}

{\fontsize{16}{18}\selectfont $hXX$}
\end{minipage}
\hfill
\begin{minipage}[t][4.00cm][c]{0.30\textwidth}
\centering
\begin{tikzpicture}[baseline=(current bounding box.center), font=\Large, line width=0.95pt]
  \coordinate (Y)  at (-1.55,0);
  \coordinate (V)  at (0,0);
  \coordinate (X1) at (1.25, 0.95);
  \coordinate (X2) at (1.25,-0.95);

  \fill (V) circle (2.0pt);

  \draw[decorate, decoration={snake, segment length=5.2pt, amplitude=1.2pt}] (Y) -- (V);
  \draw (V) -- (X1);
  \draw (V) -- (X2);

  \node[left=2pt] at (Y) {$Y$};
  \node[right=2pt] at (X1) {$X$};
  \node[right=2pt] at (X2) {$X$};
\end{tikzpicture}

\vspace*{\simplelabelshift}

{\fontsize{16}{18}\selectfont $YXX$}
\end{minipage}
\hfill
\begin{minipage}[t][4.00cm][c]{0.30\textwidth}
\centering
\begin{tikzpicture}[
  baseline=(current bounding box.center),
  x=1cm,y=0.56cm,
  font=\Large,
  line width=0.95pt,
  >=Stealth
]

\coordinate (O)       at (0.00, 0.00);

\coordinate (CpiU)    at (2.52, 1.95);
\coordinate (CrhoD)   at (2.52,-1.95);
\coordinate (CrhoU)   at (3.05, 0.70);
\coordinate (CpiD)    at (3.05,-0.70);

\coordinate (VqqGup)  at ($(O)!0.44!(CpiU)$);
\coordinate (VqqGdn)  at ($(O)!0.44!(CrhoD)$);

\coordinate (piUin)   at ($(O)!0.91!(CpiU)$);
\coordinate (rhoDin)  at ($(O)!0.91!(CrhoD)$);

\coordinate (G3up)    at (1.70, 0.40);
\coordinate (G3dn)    at (1.70,-0.40);

\coordinate (VgqqUp)  at (2.24, 0.68);
\coordinate (VgqqDn)  at (2.24,-0.68);

\coordinate (rhoUq)   at ($(CrhoU)+(-0.38, 0.14)$);
\coordinate (rhoUqb)  at ($(CrhoU)+(-0.38,-0.14)$);

\coordinate (piDq)    at ($(CpiD)+(-0.30, 0.11)$);
\coordinate (piDqb)   at ($(CpiD)+(-0.30,-0.11)$);

\fill (O)       circle (2.2pt);
\fill (VqqGup)  circle (2.0pt);
\fill (VqqGdn)  circle (2.0pt);
\fill (G3up)    circle (2.0pt);
\fill (G3dn)    circle (2.0pt);
\fill (VgqqUp)  circle (2.0pt);
\fill (VgqqDn)  circle (2.0pt);

\draw (O) -- (VqqGup) -- (piUin);
\draw (O) -- (VqqGdn) -- (rhoDin);

\coordinate (qin1)  at ($(O)!0.28!(VqqGup)$);
\coordinate (qin2)  at ($(O)!0.53!(VqqGup)$);
\draw[-{Latex[length=2.0mm]}] (qin1) -- (qin2);

\coordinate (qbin1) at ($(O)!0.28!(VqqGdn)$);
\coordinate (qbin2) at ($(O)!0.53!(VqqGdn)$);
\draw[-{Latex[length=2.0mm]}] (qbin2) -- (qbin1);

\coordinate (qout1)  at ($(VqqGup)!0.46!(piUin)$);
\coordinate (qout2)  at ($(VqqGup)!0.64!(piUin)$);
\draw[-{Latex[length=2.0mm]}] (qout1) -- (qout2);

\coordinate (qbout1) at ($(VqqGdn)!0.46!(rhoDin)$);
\coordinate (qbout2) at ($(VqqGdn)!0.64!(rhoDin)$);
\draw[-{Latex[length=2.0mm]}] (qbout2) -- (qbout1);

\node[above left=-1pt and 0pt] at ($(O)!0.60!(VqqGup)$) {$q_D$};
\node[below left=-1pt and 0pt] at ($(O)!0.60!(VqqGdn)$) {$\bar q_D$};

\draw[gluonline] (VqqGup) -- (G3up);
\draw[gluonline] (VqqGdn) -- (G3dn);
\draw[gluonline] (G3up)   -- (G3dn);
\draw[gluonline] (G3up)   -- (VgqqUp);
\draw[gluonline] (G3dn)   -- (VgqqDn);

\draw (VgqqUp) -- (rhoUq);
\draw (VgqqUp) -- (rhoUqb);

\coordinate (ru1)  at ($(VgqqUp)!0.35!(rhoUq)$);
\coordinate (ru2)  at ($(VgqqUp)!0.65!(rhoUq)$);
\draw[-{Latex[length=1.8mm]}] (ru1) -- (ru2);

\coordinate (rub1) at ($(VgqqUp)!0.35!(rhoUqb)$);
\coordinate (rub2) at ($(VgqqUp)!0.65!(rhoUqb)$);
\draw[-{Latex[length=1.8mm]}] (rub2) -- (rub1);

\draw (VgqqDn) -- (piDq);
\draw (VgqqDn) -- (piDqb);

\coordinate (pd1)  at ($(VgqqDn)!0.35!(piDq)$);
\coordinate (pd2)  at ($(VgqqDn)!0.65!(piDq)$);
\draw[-{Latex[length=1.8mm]}] (pd1) -- (pd2);

\coordinate (pdb1) at ($(VgqqDn)!0.35!(piDqb)$);
\coordinate (pdb2) at ($(VgqqDn)!0.65!(piDqb)$);
\draw[-{Latex[length=1.8mm]}] (pdb2) -- (pdb1);

\node[darkblobpi]  at (CpiU)  {$\pi_D$};
\node[darkblobrho] at (CrhoU) {$\rho_D$};
\node[darkblobpi]  at (CpiD)  {$\pi_D$};
\node[darkblobrho] at (CrhoD) {$\rho_D$};

\end{tikzpicture}

\vspace*{\darklabelshift}

{\fontsize{16}{18}\selectfont Dark $\rho$s}
\end{minipage}

\caption{Representative models featuring events with decays of pairs of particles $X$ (di-decays). Left: particles $X$ with trilinear coupling to the Higgs bosons $h$, which gives rise to decays of $B$ mesons $B_s\to XX$ and $B\to K+XX$. Middle: particles $X$ with trilinear coupling to some new short-lived resonance $Y$. Right: dark QCD models, where dark quarks $q_{D}$ (confined in dark mesons $\pi_{D},\rho_{D}$) couple to the SM via a mediator $Z'$, with $X$ being a dark meson $\rho_{D}$. These models differ significantly in particle content and underlying dynamics.}
\label{fig:representative_models}
\end{figure*}

The key advantage of di-decays lies in the kinematics information about the whole event. Namely, by reconstructing both decaying FIPs, one gains access to correlated observables such as the total invariant mass $m_{\text{inv}}$, which manifestly stores the information about the production mechanism. 

\begin{figure}[t]
    \centering
    \includegraphics[width=\linewidth]{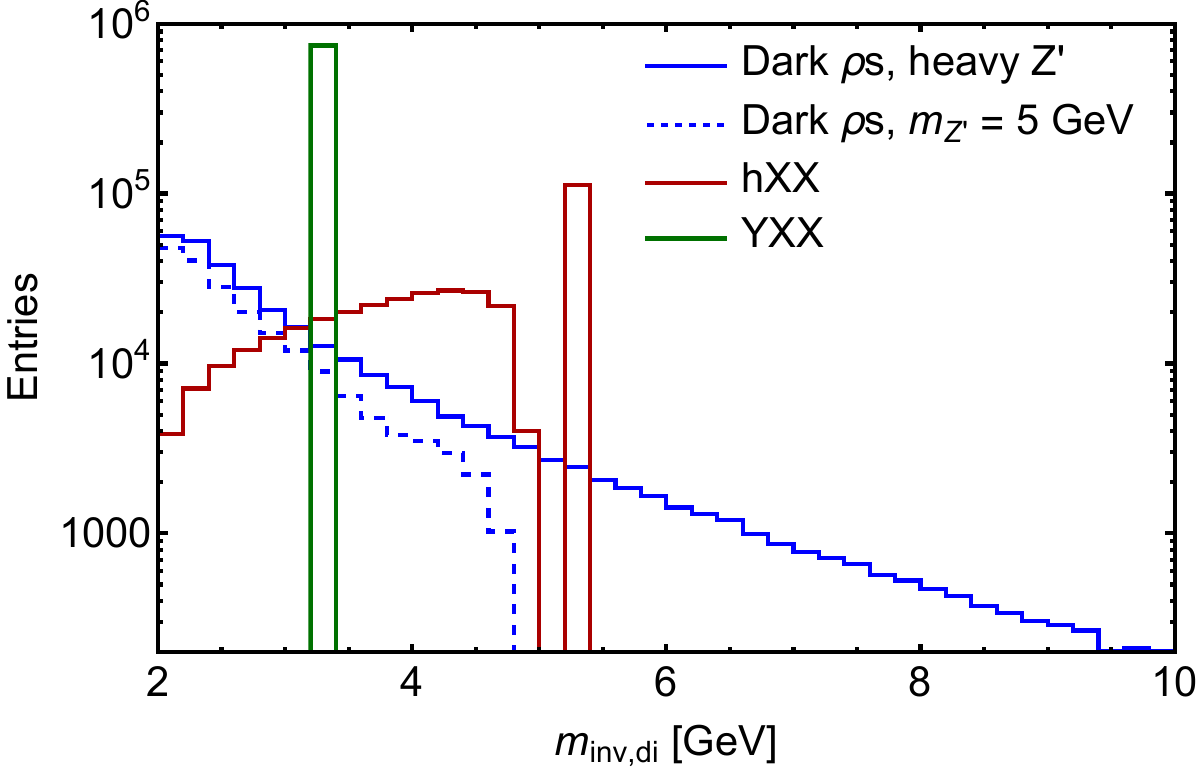}
    \caption{The MC-truth invariant mass distribution of a pair of FIPs $X$ with mass $m = 1\,\text{GeV}$ for the various models discussed in Fig.~\ref{fig:representative_models}, considering the SPS energy setup. For the dark $\rho$s model, we consider dark $\rho^{0}$ mesons in dark QCD model of Ref.~\cite{Bernreuther:2019pfb}, either with a very heavy $Z'$ boson or with $m_{Z'} = 5\,\text{GeV}$, simulated using \texttt{PYTHIA8}~\cite{Bierlich:2022pfr}. For the $YXX$ model, we fix the mass of a new resonance by $m_{Y} = 3\,\text{GeV}$. Reconstructing the invariant mass distribution from di-decay events enables straightforward identification of the production mode and underlying model, even without access to the production vertex. }
    \label{fig:invariant-masses}
\end{figure}

For instance, let us assume that an IF experiment has observed a dark-photon-like mono-decay signature. It may correspond to three entirely different FIPs: the minimal portal DP, the DP having additional quadratic coupling to the Higgs, and the dark $\rho$ mesons. The presence of di-decays would exclude the former scenario, but it is not yet enough to differentiate between the latter two models. However, we may access the $m_{\text{inv}}$ distribution, see Fig.~\ref{fig:invariant-masses}. At SPS, the $hXX$ coupling gives rise to the decays $B\to X_{s}+2X$, $B_{s} \to 2X$~\cite{Boiarska:2019jym,DallaValleGarcia:2023xhh}, with $X_{s}$ being a meson containing an $s$ quark. The $m_{\text{inv}} = \sqrt{(p_{X,1}+p_{X,2})^{2}}$ distribution is continuous for $m_{\text{inv}} < m_{B}-m_{\pi}$, and then has a sharp peak at $m_{\text{inv}} = m_{B_{s}}$. In contrast, the distribution of dark $\rho$s is maximal at small masses and may, depending on the model, continuously extend above $m_{B_{s}}$. As a result of this clear difference, it may be possible to differentiate between the modes (and hence the FIPs) by observing up to 100 events, assuming negligible backgrounds and the invariant mass resolution of $0.1\text{ GeV}$.

Other correlated observables are discussed in \cref{app:insights}.

\subsection{When are di-decays experimentally relevant?}

The principal limitation of di-decays is parametric suppression: both produced FIPs must decay inside the fiducial volume. Compared to mono-decays, the event rate is therefore suppressed by an additional factor of the decay probability. In terms of pure exclusion reach, di-decays will typically probe a smaller region of parameter space. However, this suppression must be weighed against one important advantage: substantially reduced SM backgrounds since it is far more difficult for SM processes -- whether physical or combinatorial -- to fake an event featuring two similar decay vertices. To make these statements quantitative, we consider a generic quadratic setup that captures the essential parametric features of many FIP scenarios.

Consider a FIP $X$ described by the interaction Lagrangian
\begin{equation}
    \mathcal{L}_{\text{int}} = c_{1}\, X \cdot \mathcal{O}_{1}(\Psi_{\text{SM}}) 
    + c_{2}\, \mathcal{O}_{2}(\Psi_{\text{SM}})\cdot X\cdot X\,,
    \label{eq:model-di}
\end{equation}
where $\mathcal{O}_{1,2}(\Psi_{\text{SM}})$ are operators constructed purely from SM fields $\Psi_{\text{SM}}$, and $c_{1,2}$ are independent couplings (see Eq.~\eqref{eq:lagr-scalar} for an explicit realization). The only structural assumption is that the FIP possesses both linear and quadratic couplings to the SM.

The linear operator $\mathcal{O}_1$ induces single production of $X$ and controls its decay width. Parametrically,
\begin{equation}
N_{\text{prod}}^{X} \propto c_1^2\,,
\qquad 
\tau_X \propto c_1^{-2}\,.
\end{equation}
The quadratic operator $\mathcal{O}_2$ contributes only to pair production,
\begin{equation}
    Y \to 2X,\quad Y \to Y' + 2X,\ \dots,
    \label{eq:di-production}
\end{equation}
with
\begin{equation}
N_{\text{prod}}^{2X} \propto c_2^2\,.
\end{equation}

At intensity frontier experiments, two experimental signatures may therefore arise:

\begin{itemize}
\item \emph{Mono-decays}: only one $X$ decays inside the fiducial volume (both $\mathcal{O}_1$ and $\mathcal{O}_2$ contribute to production);
\item \emph{Di-decays}: both particles decay within the same event (only $\mathcal{O}_2$ contributes).
\end{itemize}

We may utilize a simple analytical approach~\cite{Bondarenko:2019yob} to roughly estimate the number of events for these two signatures:
\begin{align}
    N_{\text{events}}^{\text{(di)}} &= N_{\text{prod}}^{2X}\times \xi^{2}, \label{eq:Nevents-di}
\\ 
    N_{\text{events}}^{\text{(mono)}} &= N_{\text{prod}}^{X+2X} \times \xi -  2N_{\text{events}}^{\text{(di)}}\,.
\end{align}
Here, explicitly, $N_{\text{prod}}^{2X} = \sum_{Y}N_{Y}\cdot \text{Br}_{Y\to 2X}$ and $N_{\text{prod}}^{X+2X} = N_{\text{prod}}^{X}+2N_{\text{prod}}^{2X}$, where $\text{Br}_{Y\to 2X}$ stays for the collective branching ratio of the modes~\eqref{eq:di-production}. Next, $\xi = \epsilon_{X}\cdot P_{\text{dec}}^{X} \cdot \epsilon_{\text{dec}} < 1$ is the suppression of the event rate due to geometric limitations, decay probability, and event selection: 
\begin{itemize}
\item[--] $\epsilon_{X}$ is the fraction of $X$\!s whose trajectories intersect the decay volume, which is displaced from the production point to reduce background rates.
\item[--] $P_{\text{dec}}^{X}$ is the $X$'s decay probability:
\begin{equation}
P_{\text{dec}}^{X} = \exp\left[-\frac{l_{\text{min}}}{c\tau_{X}\langle\gamma_{X} \beta\rangle}\right]-\exp\left[-\frac{l_{\text{max}}}{c\tau_{X}\langle\gamma_{X} \beta\rangle}\right],
\label{eq:Pdecay}
\end{equation}
with $l_{\text{min}/\text{max}}$ being the minimal and maximal distance from the collision point covered by the decay volume, and $\langle \gamma_{X} v\rangle$ the mean $X$\!s $\gamma$ factor times velocity. 
\item[--] $\epsilon_{\text{dec}}$ is the fraction of the $X$\!s' decay events that can be reconstructed; it includes the geometric part (aka the fraction of events where the trajectories of the minimal required number of decay products are within the detector) and the reconstruction part (the suppression due to the kinematic cuts and reconstruction efficiency).
\end{itemize}
The sensitivity of an experiment is determined by the signal strength $I = N_{\text{events}}/\sqrt{N_{\text{bg}}}$, where $N_{\text{bg}}$ is the background yield corresponding to the given signature (should be replaced with $1$ if the search is background-free). The ratio of the strengths for di- and mono-decay events is
\begin{equation}
\eta = \frac{I_{\text{di}}}{I_{\text{mono}}} = \frac{N_{\text{prod}}^{2X}}{N_{\text{prod}}^{X}+2N_{\text{prod}}^{2X}(1-\xi)}\times \xi \times \sqrt{\frac{N_{\text{bg}}^{\text{mono}}}{N_{\text{bg}}^{\text{di}}}},
\label{eq:ratio-di-mono}
\end{equation} 

\begin{figure}[t]
    \centering
    \includegraphics[width=0.9\linewidth]{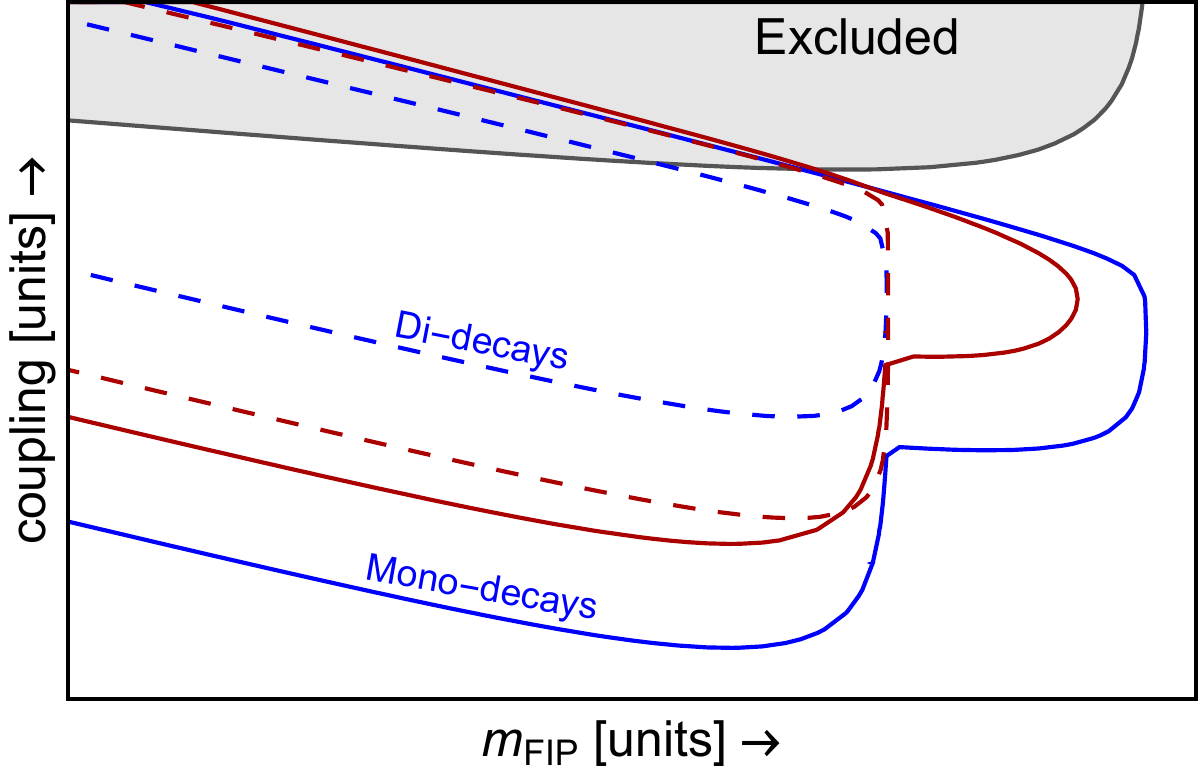}
    \caption{Schematic parameter space of a FIP having mass $m_{\text{FIP}}$ and coupling to the SM particles $g$, which may be produced at accelerator experiments in two ways -- as a single particle or in pairs. At an intensity frontier experiment, they may be probed by searching for either a single decay vertex (``mono-decay'') or two decay vertices within the same event (``di-decay''). The excluded parameter space is shown in gray, whereas the possible sensitivities of intensity frontier experiments to mono- and di-decays are indicated, correspondingly, by the solid and dashed curves. Depending on the background assumptions, geometric setup of the experiment, and FIP's kinematics, the di-decays may either compete with the mono-decay signature (the red curves) or be sub-dominant, still offering unique insights about FIPs in case of detection (the blue curves).}
    \label{fig:illustration}
\end{figure}

There is a non-trivial interplay between the different factors in Eq.~\eqref{eq:ratio-di-mono} determining the relative coverage of the parameter space by the signatures; see Fig.~\ref{fig:illustration}. 

Provided that $N_{\text{prod}}^{2X}/N_{\text{prod}}^{X}$ is not suppressed, $\eta$ critically depends on how significantly $\xi$ deviates from its maximal value 1. At high-energy beam-dump or collider experiments with limited angular coverage of detectors and large FIP's boosts~\cite{SHiP:2015vad,NA62:2023qyn,FASER:2018bac,Gorkavenko:2023nbk}, $\xi \ll 1$, so di-decays typically contribute only at larger couplings, i.e., for shorter lifetimes. Still, it may be non-negligible and deliver the opportunities that we discuss below. 

The $\xi$ suppression is milder when FIPs have modest boosts and the detector subtends a large fraction of the solid angle, as at Belle~II~\cite{Belle-II:2018jsg}. Future $e^{+}e^{-}$ facilities such as FCC-ee~\cite{Blondel:2022qqo} provide interesting settings in which to study the same effect.

The background factor in Eq.~\eqref{eq:ratio-di-mono} becomes important whenever the mono-decay search is not background-free, as for Downstream@LHCb~\cite{Gorkavenko:2023nbk} and Belle~II~\cite{Belle-II:2023ueh}. Proposed off-axis experiments such as ANUBIS~\cite{Bauer:2019vqk} provide additional settings where the background rejection from two correlated vertices may be useful. While the mono-decay signature may be contaminated by multiple backgrounds, di-decay events are much cleaner: reconstructing two displaced vertices with correlated decay products is significantly harder to fake combinatorially.

Thus, in settings with moderate boosts and/or non-negligible backgrounds, di-decays can substantially surpass or complement standard single-decay searches. Moreover, beyond pure sensitivity, di-decays provide direct access to the production mechanism through their correlation structure and multiplicity. In scenarios where both $c_1$ and $c_2$ are present, mono-decays alone cannot disentangle whether the observed rate arises predominantly from linear or quadratic interactions. The observation (or absence) of di-decays lifts this degeneracy and constrains the underlying operator structure.

In the next section, we make these considerations concrete in a minimal renormalizable setup -- a singlet scalar coupled to the Higgs sector -- where linear and quadratic operators arise naturally, and di-decays emerge as a direct probe of the Higgs portal structure.

\section{Higgs-like scalars}
\label{sec:scalar-model}

We consider a minimal singlet-scalar extension of the SM. We introduce a single scalar $S$ with the generic renormalizable Lagrangian 
\begin{equation}
 \mathcal{L}\supset  \Lambda_S SH^{\dagger}H + \lambda_s S^2 H^{\dagger}H \, ,
 \label{eq:lagr-scalar}
\end{equation} where $H$ is the SM Higgs doublet.
Below the electroweak symmetry-breaking scale, it is effectively reduced to~\cite{Boiarska:2019vid} 
\begin{equation}
    \mathcal{L} \supset m_{h}^{2}\theta h S + \frac{\alpha}{2}h SS +\cdots\, ,
    \label{eq:lagr-eff}
\end{equation} 
where $h$ is the SM Higgs boson with mass  $m_h \approx125$~GeV, $\theta = v \Lambda_S /m_h^2$  and $\alpha = 2v \lambda_s$ with $v=246$~GeV the electroweak vacuum expectation value. The first term describes the mixing angle, whereas the second one is the trilinear coupling. The coupling $\alpha$ can be expressed  in terms of the fixed branching ratio $\text{Br}_{h\to SS}$ as
\begin{equation}
    \alpha^{2} \approx \frac{32\pi }{\sqrt{1-\frac{4m_{S}^{2}}{m_{h}^{2}}}}\text{Br}_{h\to SS}\Gamma_{h}m_{h} \,.
    \label{eq:alpha-Br-h-SS}
\end{equation}

The model is attractive for several reasons. First, it is well motivated, as it is simple and yet may be connected to inflation or be a mediator between dark matter and SM~\cite{Bezrukov:2009yw,Fradette:2018hhl,Monin:2018lee,Winkler:2018qyg,Krnjaic:2015mbs}. Hence, it is widely considered by the experimental community~\cite{Beacham:2019nyx,Antel:2023hkf,Ferber:2023iso,AlemanyFernandez:2927631,Belle-II:2023ueh,SHiP:2025ows}. Second, similarly to other models with FIPs, only the mono-decay signature has been previously studied at the IF experiments.

Phenomenologically, the $S$ particles are ``light Higgs bosons'', with the interactions suppressed by the small mixing angle $\theta\ll 1$ and an additional trilinear $hSS$ coupling. The main constraints on the $\alpha$ come from unobserving the decays of the type $h\to \text{inv}$, setting the model-independent bound $
\text{Br}(h\to \text{inv})<0.15$ at 95\% CL~\cite{CMS:2023sdw},\footnote{With the HL-LHC runs, the sensitivity to invisible Higgs decays will be improved down to a few percent~\cite{Bernaciak:2014pna,Bechtle:2014ewa}.} and the events $pp \to h \to 2X \to 4\text{SM}$, where $\text{SM} = \mu, b,\gamma$~\cite{CMS:2012qms,ATLAS:2023ian,CMS:2024jyb,CMS:2024vjn,CMS:2024zfv}. In particular, Ref.~\cite{CMS:2024jyb} sets the CMS constraint on the dimuon decay channel for the particles having the proper lifetimes $c\tau_{X}<100\text{ mm}$. 

It is worth mentioning the common case for a complex dark scalar singlet $S \neq S^\ast$ acquiring a vacuum expectation value $w$ (i.e., $S=(s+w)/\sqrt{2}$ in the unitary gauge), for which $\Lambda_S = 2\lambda_s w $. In this case, we have $\theta m_h/\alpha = w/m_h \sim m_S/m_h$. Thus, one naturally expects tiny mixing $\theta \lesssim 10^{-4}$  for small branching ratios  $\text{Br}_{h\to SS}\sim0.01$ ($\alpha\sim1$~GeV) and light scalar masses $m_S < m_{B}/2$ -- the region of parameter space we will be interested.\footnote{A $Z_2$ symmetry where only the dark scalar transforms as $S \;\to\; -S$ would also motivate tiny mixing angles for small values of  $\text{Br}_{h\to SS}$ in similar way. The term $\Lambda_S S H^\dagger H$ is odd and can be naturally suppressed, while the quartic portal $\lambda_s S^2 H^\dagger H$ is $\mathbb{Z}_2$-even and can remain unsuppressed. }

In \cref{app:scalar-pheno}, we discuss further details on the model. In particular, the scalar decays into SM particles, taking into account theoretical uncertainties that are highly relevant in the GeV mass regime.

\subsection{Higgs-like Scalar Production}
\label{sec:prod-decay}

The effective interactions~\eqref{eq:lagr-eff} generate the following production modes~\cite{Wilczek:1977zn,Boiarska:2019jym}: single scalars,
\begin{equation}
   B^{+/0}\to Y_{s/d}+S, \quad B_{s}\to \phi + S,  \quad  Y_{V} \to \gamma+S,
    \label{eq:mono-production-scalar}
\end{equation}
and scalar pairs,
\begin{align}
\label{eq:di-production-scalar}
&h \to SS, \quad B_{s}\to SS, \quad B^{+/0} \to Y_{s/d}+SS, \\ &B_{s}\to \phi + SS, \quad  Y_{V} \to \gamma+SS \, 
    \label{eq:di-production-radiative}
\end{align}
Here, $Y_{s/d}$ is a hadronic state containing an $s/d$ quark: $\pi, K, K^{*}(892,\dots),K_{1}(1270,\dots),K_{2}^{*},K_{0}^{*}$, while $Y_{V}$ is an arbitrary vector meson (the most interesting are the cases of $\Upsilon$ and $J/\psi$ mesons). The mixing coupling $\theta hS$ induces the first group of processes, whereas the second and third groups originate from the $hSS$ interaction. Decays of $B$ mesons are mediated by the flavor-changing neutral currents (FCNC) $b\to s/d$ effectively generated by electroweak loops with a $W$ boson and a $t$ quark, with an outgoing scalar or a pair of scalars. Here, we neglect the processes of the proton bremsstrahlung and decays of kaons, which are mostly subdominant in the SPS- and LHC-based experiments we are considering, although they may also produce pairs of scalars.

The production channels~\eqref{eq:mono-production-scalar},~\eqref{eq:di-production-scalar} have been studied in detail in Ref.~\cite{Boiarska:2019jym}. For the decays of $B$ mesons via the mixing, summing over all possible final states and in the limit $m_{S}\ll m_{B}$, we get $\text{Br}(B\to S +Y)\approx 3.3\cdot \theta^{2}$. Considering the decays of $B$ mesons into two scalars under the same limit, we get $\text{Br}_{B\to SS+Y} \approx 10^{-9}\left(\frac{\alpha}{1\text{ GeV}}\right)^{2}$. The probability of the radiative production processes $Y_{V}\to \gamma+S/SS$, not considered in Ref.~\cite{Boiarska:2019jym}, is too small for this process to contribute to the event yield, see \cref{app:scalar-pheno}.

The production probabilities from these channels (per proton collision) are \begin{equation}
    P_{\text{prod},S}^{(Y)} = \chi_{Y}\cdot \text{Br}_{Y\to S} \,,
\end{equation} where $\chi_{Y}\equiv \sigma_{pp\to Y}/\sigma_{pp,\text{total}}$ is the yield of the mother particles per proton collision. For the $B$ mesons, we define 
\begin{equation}
\sigma_{pp\to B} = 2\cdot \sigma_{pp\to bb}\cdot \kappa_{b\bar{b}}\cdot f_{b\to B},
\end{equation}
with a factor of $2$ staying for including $B$s and anti-$B$s, $\sigma_{pp}$ for the $b\bar{b}$ production cross-section, $\kappa_{b\bar{b}}$ for the cascade enhancement of the $b\bar{b}$ production in the case of the finite thickness target, whereas $f_{b\to B}$ for the fragmentation fraction.

\begin{figure}[t]
    \centering
    \includegraphics[width=0.9\linewidth]{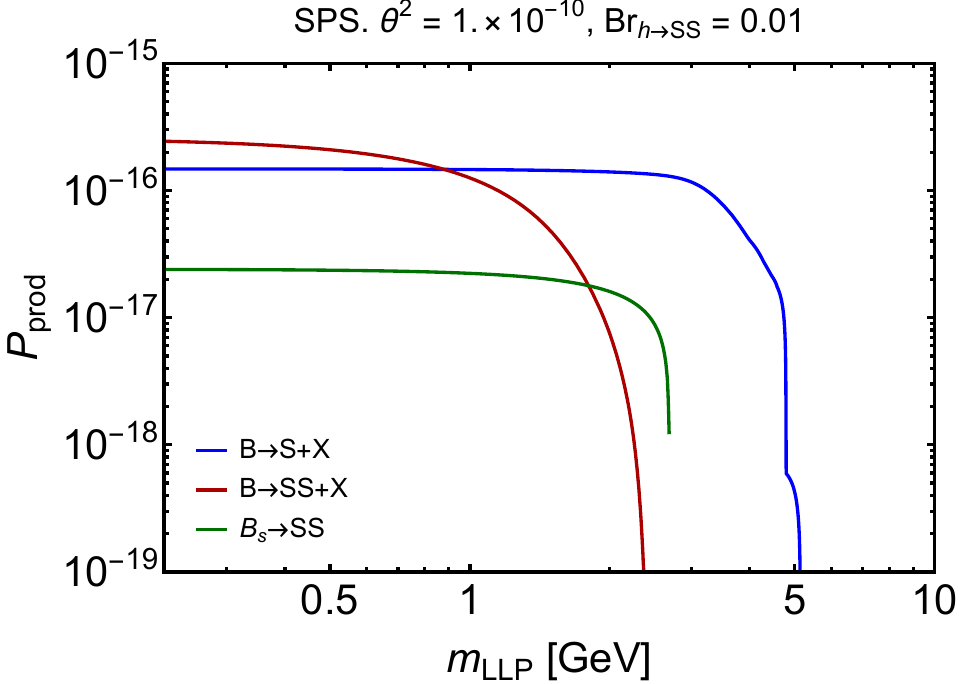}\\\includegraphics[width=0.9\linewidth]{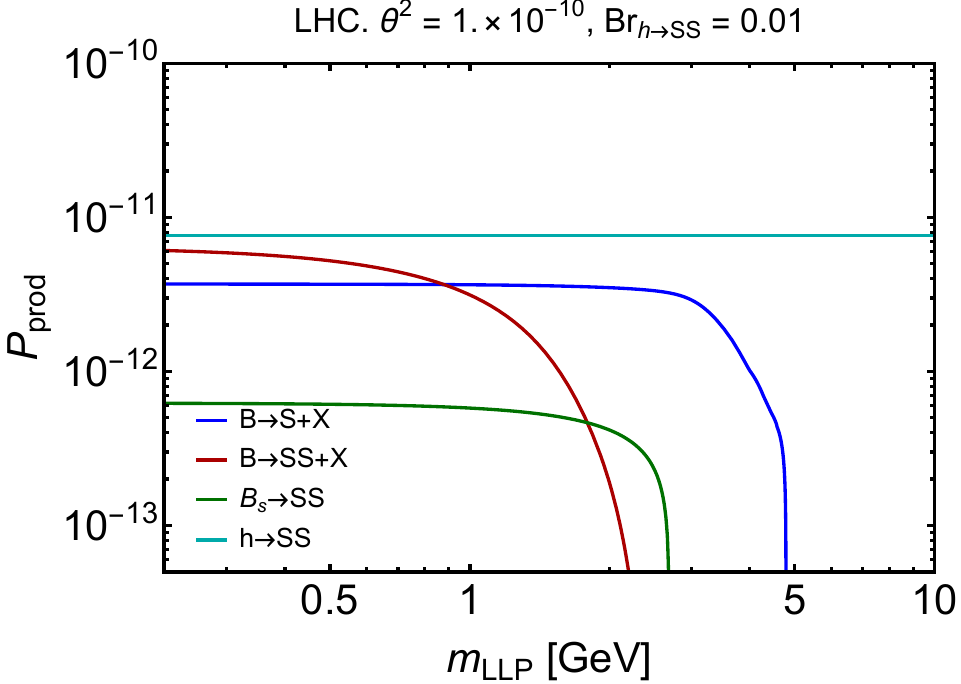}
    \caption{Production probabilities of Higgs-like scalars for particular values of $\theta^{2}$ and $\text{Br}_{h\to SS}$ at SPS and LHC facilities. The figures have been produced using \texttt{SensCalc}~\cite{Ovchynnikov:2023cry,SensCalc-Zenodo}. We do not show the production from radiative decays of $\Upsilon$ and $J/\psi$ mesons as it leads to a negligible contribution to the number of events, see Sec.~\ref{app:vector-decays}.}
    \label{fig:production-probabilities}
\end{figure}

The plot of the production probabilities for SPS and LHC facilities assuming the value $\text{Br}_{h\to SS} = 0.01$ is shown in Fig.~\ref{fig:production-probabilities}. 

At the LHC (where the $h\to SS$ mode is accessible), the modes~\eqref{eq:di-production-scalar} dominate the whole scalar production provided that $\theta^{2}/\text{Br}(h\to SS) \lesssim 2\cdot 10^{-8}$. At the facilities where $h$ cannot be produced, the di-production processes are the main mechanisms under the same condition, but only for masses $m_{S}\lesssim m_{B_{s}}/2$~\cite{Boiarska:2019vid}. Provided that $\text{Br}(h\to SS)\gtrsim 10^{-4}$, these couplings are reachable by the Intensity Frontier experiments~\cite{Beacham:2019nyx,Aberle:2839677}.
 
\section{Experimental set up}

To study the potential of IF experiments to explore scalars using di-decays, we consider three distinct setups -- SHiP~\cite{SHiP:2025ows}, Belle II~\cite{Belle-II:2018jsg}, and the Downstream algorithm at LHCb~\cite{Gorkavenko:2023nbk,Kholoimov:2025cqe} (which we will call Downstream@LHCb throughout the text). The latter two are currently running, while the former has been approved in 2024 and is expected to start operating in 2033. Further details of our analysis, including the description of the accurate simulation we conducted to perform the study and the experimental setups, are summarized in Appendices~\ref{app:experiments} and~\ref{app:event-sampler}.

SHiP is a proton beam dump experiment with the beam energy $E_{p} = 400\text{ GeV}$ and a projective decay volume separated by a macroscopic distance from the target. Detailed simulations show that the searches for decays of FIPs at SHiP will operate in the background-free regime during the full running time of the experiment, which is 15 years~\cite{Aberle:2839677}.

Downstream@LHCb is a recently proposed trigger scheme (already implemented in the LHCb system and collecting data) that enables searching for decay events outside the inner LHCb tracker and up to the UT tracker. It allows for the simultaneous enlargement of the effective decay volume beyond the inner LHCb tracker and reduces backgrounds, complementing the prompt LHCb searches with the algorithm to explore the parameter space of FIPs. The domain of invariant masses $m>2\text{ GeV}$ is expected to be background-free independently of the decay event topology, thanks to the absence of long-lived physical resonances above this mass and a huge suppression of combinatorial events~\cite{Kholoimov:2025cqe,LHCB-FIGURE-2025-002}. The background reduction study for smaller masses is in progress. The search may continue until the high luminosity phase of the LHC, with the total luminosity $\mathcal{L} = 300\text{ fb}^{-1}$.

At Belle II, collisions of electrons and positrons produce pairs of $B\bar{B}$ particles with small boosts. Their decays may be used to search for FIPs within the tracking volume surrounding the interaction point. Unlike SHiP and Downstream@LHCb, Belle II has access to the production vertex, at a cost of a significant background. Namely, for the chain $B^{+}\to K^{+}X, \ X \to \text{visible}$, which is the main search strategy for Higgs-like particles and ALPs at Belle II~\cite{Belle-II:2018jsg}, the background is at the level of $5-10$ events for the luminosity $\mathcal{L} = 189\text{ fb}^{-1}$~\cite{Belle-II:2023ueh}. Assuming the luminosity of the full run $\mathcal{L} =50\text{ ab}^{-1}$, it scales to $(1-2)\cdot 10^{3}$.

The three setups probe complementary kinematic and experimental regimes:
\begin{itemize}
    \item[--] \textit{SHiP} combines an exceptionally large heavy-meson yield, a 50~m decay volume, and a dedicated low-background design. Its restricted forward aperture and macroscopic baseline limit the geometric and decay acceptance, while the large scalar boosts increase the laboratory decay length at fixed proper lifetime. The same forward boost, however, collimates the scalar pair and favors the simultaneous geometric acceptance of both particles.
    \item[--] \textit{Belle~II} offers broad angular coverage around the interaction point and relatively small $B$-meson boosts, both of which favor the acceptance of two displaced decays. In the channels $B\to K^{(*)}S$ and $B\to K^{(*)}SS$, the prompt kaon also permits reconstruction of the production chain. These advantages are balanced by the smaller number of produced $B$ mesons, the finite displaced-tracking volume, and the non-negligible backgrounds affecting mono-decay searches.
    \item[--] \textit{Downstream@LHCb} benefits from the large LHC heavy-flavor yield and, through $h\to SS$, accesses scalar masses up to $m_h/2$. Its short forward fiducial region and the large scalar boosts reduce the decay probability, while neutral or soft final-state particles can lead to partial reconstruction. We therefore restrict the projected sensitivity to the selected region $m_{\rm inv}>2$~GeV, where the background is expected to be negligible.
\end{itemize}
Let us illustrate the relevance of di-decays considering SHiP as an example. The forward boost of the produced scalars gives a sizeable correlated pair acceptance. For each scalar, define the geometric factor
\begin{equation*}
    g_i \equiv \epsilon_{S,i}\,\epsilon_{{\rm dec},i}^{\rm geom},
\end{equation*}
where $\epsilon_{S,i}$ requires the scalar trajectory to intersect the decay volume and $\epsilon_{{\rm dec},i}^{\rm geom}$ requires the selected visible decay products to point toward the detector. For SHiP, a representative value for the product is $\langle g_i\rangle\simeq0.2$~\cite{Ovchynnikov:2023cry}. Neglecting correlations only for this order-of-magnitude illustration,\footnote{The full calculation does not make this factorization approximation: it evaluates the correlated quantity $A_{\rm geom}^{(2)}=\langle g_1g_2\rangle$ event by event, using the joint kinematics of the scalar pair.} the joint geometric acceptance is therefore
\begin{equation*}
    A_{\rm geom}^{(2)}\simeq \langle g_i\rangle^2\simeq 0.04.
\end{equation*}
Thus, the requirement of accepting both scalar trajectories and both visible decays retains a percent-level fraction of the produced pairs. Reconstruction is not included in the definition of $g_i$; for the selected charged topologies, its efficiency is assumed to be of order unity.

The decay probability supplies an additional, conceptually separate suppression. Defining the longitudinal laboratory decay length $\lambda_z=c\tau_S\beta\gamma\cos\vartheta$, where $\vartheta$ is the polar angle relative to the beam axis, the SHiP fiducial interval gives
\begin{equation*}
    P_{\rm dec}(\lambda_z)
    =e^{-32\,{\rm m}/\lambda_z}-e^{-82\,{\rm m}/\lambda_z}.
\end{equation*}
It is maximal at $\lambda_z=(82-32)\,{\rm m}/\ln(82/32)\simeq53\,{\rm m}$, where $P_{\rm dec}^{\rm max}\simeq0.33$. Consequently, close to the optimum, a transparent factorized estimate of the combined geometric and double-decay probability is
\begin{equation*}
    \left(0.2\,P_{\rm dec}^{\rm max}\right)^2\simeq4\times10^{-3}.
\end{equation*}
This estimate isolates the geometric and decay-in-volume factors; visible branching fractions and reconstruction enter separately in the numerical event rate. The resulting percent-to-per-mille efficiency remains phenomenologically relevant because of the very large number of heavy mesons produced at SHiP. In the long-lifetime regime, the corresponding single-scalar probability approaches
\begin{equation*}
    \left\langle P_{\rm dec}\right\rangle
    \simeq \frac{50\,{\rm m}}{c\tau_S}
    \left\langle\frac{1}{\beta\gamma\cos\vartheta}\right\rangle,
\end{equation*}
which controls the lower-coupling boundary of the projected sensitivity. 

Accidental overlaps can, in principle, fake a di-decay signature in two ways: through coincidence of two independent mono-decays produced in beam-induced collisions, or through overlap of a genuine mono-decay with unrelated cosmic activity. However, as we quantify in Appendix~\ref{app:pair-coincidences} for SHiP, the expected rate of beam-induced fake di-decays is negligible in the parameter space relevant for our study. The same conclusion applies even more strongly to Belle~II and LHCb, owing respectively to the essentially pileup-free environment at Belle~II and to the shorter effective decay region and bunch-crossing timing structure at LHCb. Cosmic-induced overlaps are likewise strongly suppressed by timing, spatial reconstruction, and momentum directionality; for the scalar case considered here, the additional requirement that the two vertices reconstruct a consistent common invariant mass provides a further rejection handle.

\section{Results} 

\begin{figure}[t]
    \centering
    \includegraphics[width=\linewidth]{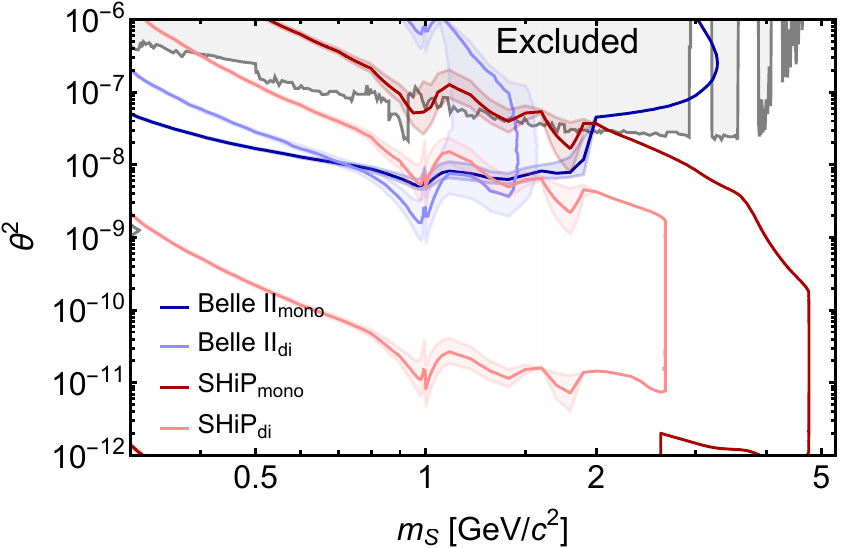}\\ \includegraphics[width=\linewidth]{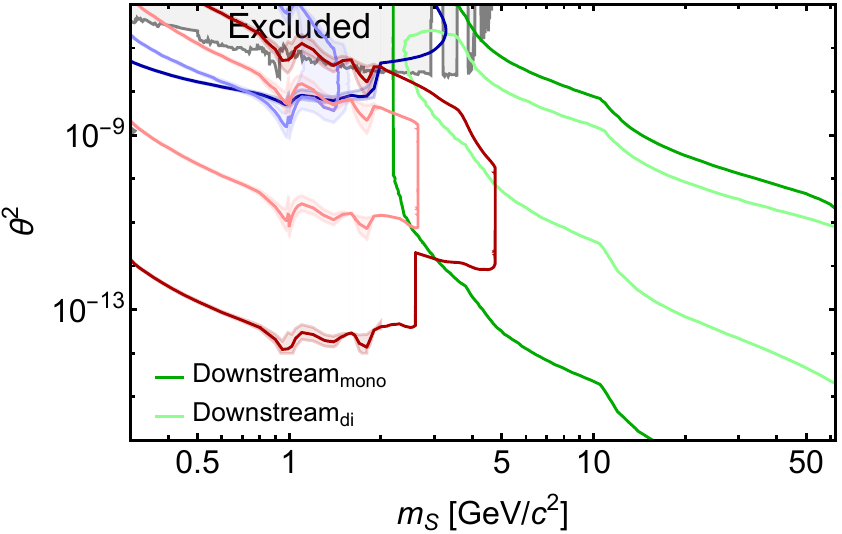}
    \caption{Parameter space of Higgs-like scalars in the plane mass-coupling to be probed with mono- and di-decay signatures at SHiP, Belle II, and Downstream@LHCb. The mono-decay domain constitutes the exclusion potential, while the di-decay region is where it may be possible to differentiate between various models with the same Higgs-like particle decay pattern. The coupling $\alpha$ in Eq.~\eqref{eq:lagr-scalar} is fixed by setting $\text{Br}_{h \to SS} = 0.1$. The top panel shows the zoomed-in parameter space probed by SHiP and Belle II, whereas the bottom panel also includes the domain to be explored by Downstream@LHCb. The excluded domain is taken from Ref.~\cite{Belle-II:2023ueh}. The shaded colored regions in the sensitivities show the uncertainties in the scalar decay description (see Appendix~\ref{app:scalar-pheno} and Ref.~\cite{Blackstone:2024ouf}).}
    \label{fig:results}
\end{figure}

In the top panel of Fig.~\ref{fig:results}, we show the sensitivity of these experiments to the traditional mono-decay signature with Higgs-like scalars, assuming full running times. For the Downstream@LHCb, we conservatively restrict ourselves to the events with the invariant mass $m_{\text{inv}}>2\text{ GeV}$.

At the domain of small couplings and in the mass range $m \lesssim m_{B}$, SHiP has the best sensitivity because of a larger number of $B$ mesons and decay volume, and zero background. On the other hand, Belle II offers a unique opportunity to reconstruct the full event chain even for mono-decays, whereas Downstream@LHCb may probe scalars produced by decays of Higgs bosons, thus covering the mass range $m_{S}\gtrsim m_{B}$.

Let us now discuss the di-decay signature. The sensitivity of the experiments to the di-decays of scalars is summarized in the bottom panel of Fig.~\ref{fig:results}.

The domains of masses and coupling accessible by the three experiments nicely complement each other. Belle II and SHiP may probe masses $m < m_{B_{s}}/2$ in the different ranges of the mixing angle, whereas Downstream@LHCb accesses larger masses.

For SHiP, the di-decay sensitivity lies inside the mono-decay domain. This is mainly because of the relative suppression of the di-decay rate by the additional decay probability of the second scalar. Nevertheless, it covers a significant domain in the parameter space, where it may be possible to properly identify the production modes and robustly differentiate the model (see Appendix~\ref{app:event-sampler-SHiP} for further details). 

For Belle II, we first need to specify the background rate. The mono-decay background is fully dominated by the events with fake tracks and tracks with missing hits, not by the physical events~\cite{Belle-II:2023ueh}, and we expect the same for di-decays. The probability of forming two displaced vertices that have the same invariant mass, and would recover the $B$ meson invariant mass if combined with the kaon from the production vertex, is combinatorially suppressed. Hence, we optimistically assume that for the modes $B\to K +SS$ with decaying $S$ pair and tagged kaon $K$, the di-decay signature at Belle II is background-free, although detailed studies by the collaboration are needed to confirm this.

Under this optimistic background-free assumption, the relatively mild decay-probability suppression at Belle~II makes di-decays much more competitive than mono-decays. For scalar masses around 1~GeV, the di-decay signature can provide the leading projected sensitivity.

Finally, for Downstream@LHCb, the situation is similar to SHiP: the di-decay domain is inside the mono-decay sensitivity. But the mass range to be probed by di-decays extends until $m_{S} = m_{h}/2$, where the decay events have a very high multiplicity of the final states. Such high-multiplicity topologies are natural targets for the new Buffer Scanner (BuSca) project at LHCb~\cite{LHCB-FIGURE-2024-036,LHCB-FIGURE-2025-002}, which monitors Downstream candidates in real time at 30~MHz and could provide a dedicated route to studying high-mass di-decay signals.

Proposed off-axis LHC detectors such as MATHUSLA, ANUBIS, and CODEX-b provide natural targets for future di-decay studies~\cite{AlemanyFernandez:2927631}. Their comparatively small FIP boosts and potentially non-negligible backgrounds make the interplay among double-decay probability, pair acceptance, and two-vertex background rejection worth a dedicated study.

\section{Conclusions and outlook}
\label{sec:conclusions}

In this work, we introduced and quantified the \emph{di-decay} signature at Intensity Frontier experiments: the observation of the simultaneous decays of two Feebly Interacting Particles (FIPs) within a single event. While most existing searches focus on mono-decays, we have shown that di-decays constitute a qualitatively distinct and phenomenologically powerful probe of FIPs.

Di-decays offer three primary advantages.
First, they serve as a smoking-gun signal for pair-production mechanisms and therefore directly test an important structural feature of many FIP models.
Second, reconstructing both displaced decays provides access to correlated observables—most notably the invariant mass of the FIP pair, $m_{\rm inv}$—that encode information about the production dynamics even when the production vertex itself is experimentally inaccessible, as is typical at the Intensity Frontier.
This makes di-decays a particularly robust tool for probing the underlying operator structure and for distinguishing between classes of models that share identical decay phenomenology.
Third, the requirement of two displaced decays leads to significantly reduced background rates compared to mono-decays.
In realistic high-background environments, this cleanliness can compensate for the parametric suppression of the di-decay rate and, in some cases, yield comparable or even superior sensitivity.

Using the minimal singlet-scalar (Higgs-like scalar) extension of the Standard Model as a concrete benchmark, we demonstrated that di-decays can probe a substantial and phenomenologically relevant region of parameter space at current and upcoming Intensity Frontier experiments.
Our analysis highlights in particular the strong potential of Belle II, SHiP, and Downstream@LHCb to explore di-decay signals, offering a path to refine and expand the FIP search strategies at these facilities. Di-decays stand out as an especially promising signature for Belle~II, where mono-decay searches are background-limited. If the tagged two-vertex requirement suppresses the residual background to a negligible level, the di-decay reach can become competitive or leading.

Although our quantitative analysis has focused on the minimal singlet-scalar scenario, the strategy we have developed is broadly applicable. Any framework predicting pair-produced long-lived states—including models with additional mediators or richer hidden-sector dynamics—can give rise to di-decay signatures. In this work, we have established the phenomenological relevance and experimental competitiveness of this signature across representative Intensity Frontier environments, demonstrating that it provides structurally new information beyond mono-decays. In the discovery era, di-decays offer a concrete pathway toward genuine model discrimination by exploiting the correlated observables. A systematic exploration of this distinguishability program, including additional model classes such as dark QCD-like sectors~\cite{Bernreuther:2025xqk} or short-lived resonance structures~\cite{DallaValleGarcia:2025giv}, will be pursued in future work.

\textbf{Acknowledgements.} The authors thank Torben Ferber for discussing the setup and search analyses of Belle II, Lesya Shchutska for discussing the di-decay signature at the LHC, and Arantza Oyanguren for discussing the Downstream algorithm at LHCb. They also thank Miguel Escudero, Oleksii Mikulenko, Vsevolod Syvolap, Arantza Oyanguren, and Jonas Matuszak for reading the manuscript and providing useful comments. GG thanks the Doctoral School ``Karlsruhe School of Elementary and Astroparticle Physics: Science and Technology (KSETA)'' for financial support through the GSSP program of the German Academic Exchange Service (DAAD). GG has received support from the European Union’s Framework Programme for Research and Innovation Horizon 2020 under grant H2020-MSCA-ITN-2019/860881-HIDDeN.

\bibliography{main}

\onecolumngrid 

\appendix

\newpage

\section*{Appendix}

In this Appendix, we describe the phenomenology of Higgs-like scalars as well as the approach we use to derive our main results -- the sensitivity of SHiP, Belle II, and Downstream@LHCb experiments to di-decays.

It is organized as follows.  In Sec.~\ref{app:scalar-pheno}, we discuss some details of the phenomenology of Higgs-like scalars, including the uncertainties in the description of their decays, which, for the di-decay signature, enter the results quadratically via the squared decay probability, depending on the parameter space. 

Sec.~\ref{app:experiments} summarizes the experimental setups of SHiP, Belle~II, and the Downstream algorithm at LHCb, including the backgrounds and selection criteria used in our sensitivity estimates. In Sec.~\ref{app:pair-coincidences}, we quantify the impact of the experimental time resolution on potential backgrounds arising from accidental coincidences of independent mono-decays. Finally, Sec.~\ref{app:event-sampler} describes our experiment-agnostic framework for computing the di-decay event rate across different experimental configurations.

In Sec.~\ref{app:other-models}, we discuss whether trilinear $hXX$ coupling may be used to search for di-decays of other FIPs, such as dark photons and axion-like particles, at the facilities where Higgs bosons cannot be produced.

Finally, Sec.~\ref{app:insights} discusses various insights about di-decay events and opportunities in reconstructing the FIP's properties, using Higgs-like scalars at SHiP and the Downstream algorithm at LHCb as an example.

\section{Phenomenology of Higgs-like scalars}
\label{app:scalar-pheno}

\subsection{Decays}

Scalars decay similarly to a light Higgs boson. Therefore, they prefer decaying into the heaviest particles available by the kinematics. In particular, for decays into fundamental fermions, the decay width scales as $\Gamma_{S\to f\bar{f}}\propto y_{f}^{2}$, where $y_{f}$ is the Yukawa coupling. An exception is the decay into two gluons, which is a loop-induced process receiving contributions from heavy quarks.

The main theoretical uncertainty in the decays originates from the poor knowledge of meson spectroscopy in the sector of scalar mesons. They mix with the scalars and may severely affect the decays of $S$ into, e.g., a pair of pions and kaons. Various studies~\cite{Donoghue:1990xh,Monin:2018lee,Winkler:2018qyg,Blackstone:2024ouf} estimated the decay width of $S$s. We will follow the latest study~\cite{Blackstone:2024ouf}, which utilized the experimental data on the scatterings $\pi \pi \to KK$ and $\pi \pi \to \pi \pi$ to extract the form-factors mediating the scalar decay, and estimated the theoretical uncertainties.

To match the exclusive widths $S\to \pi \pi, KK, \dots$ with the perturbative QCD calculations including $S\to GG, s\bar{s}$ decays, we follow the approach of~\cite{Winkler:2018qyg} and add a fictitious decay width $S\to 4\pi$, fixed in a way such that the exclusive and perturbative calculations match at $m_{S}  =2\text{ GeV}$. 

\begin{figure}[h!]
    \centering
    \includegraphics[width=0.5\linewidth]{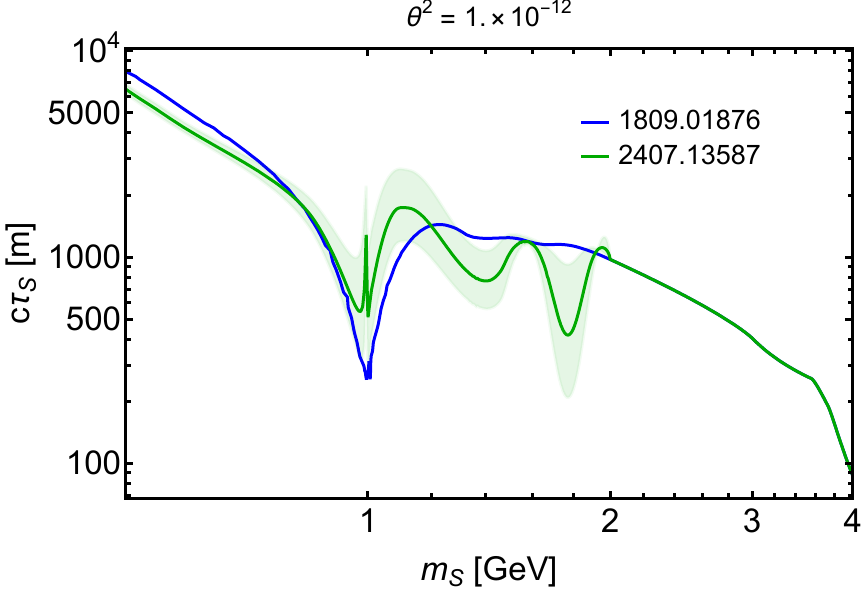}~\includegraphics[width=0.5\linewidth]{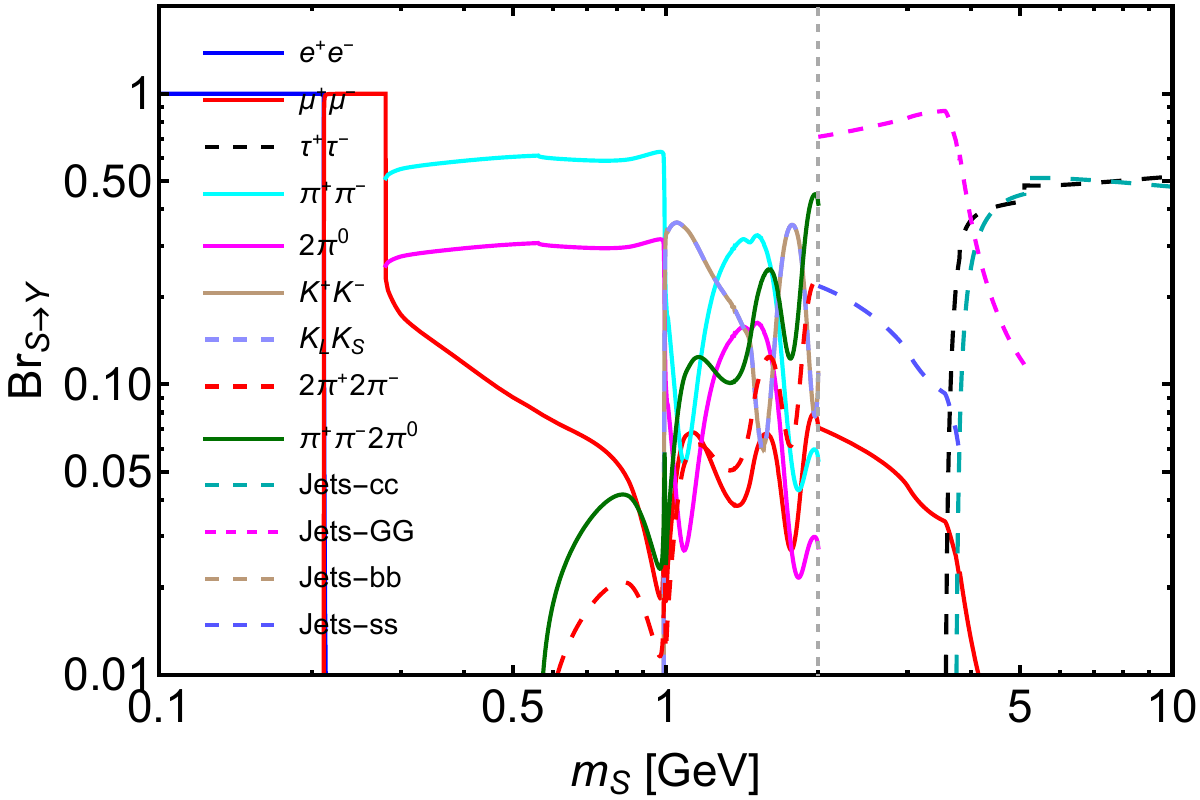}
    \caption{Phenomenology of decays of Higgs-like scalars. Left panel: behavior of the scalar lifetime $c\tau_{S}$ with the scalar mass, assuming the fixed value of the mixing angle $\theta^{2} = 10^{-12}$. The curves show two different computations of the decay width into a pair of pions and kaons: from Ref.~\cite{Blackstone:2024ouf}, including the uncertainty band (the green curve), and from Ref.~\cite{Winkler:2018qyg} (the blue curve). Right panel: the branching ratios of decays of Higgs-like scalars into various final states, assuming that the decay widths of the scalar into pairs of pions and kaons match the central prediction of Ref.~\cite{Blackstone:2024ouf}.}
    \label{fig:scalar-lifetime}
\end{figure}

The behavior of the scalar's proper lifetime $c\tau_{S}$ calculated using the $S\to \pi\pi,KK$ calculations from~\cite{Blackstone:2024ouf} is shown in Fig.~\ref{fig:scalar-lifetime} (left panel), including the uncertainties; we will utilize this description throughout the paper. For comparison, we also show the prediction of Ref.~\cite{Winkler:2018qyg}. In addition, the right panel of this figure shows the behavior of branching ratios of various scalar decay modes.

\subsection{Production in radiative decays of vector mesons}
\label{app:vector-decays}
We will concentrate on heavy mesons $V = J/\psi,\Upsilon$ in the process~\eqref{eq:di-production-radiative} due to their weaker Yukawa suppression and because we are interested in $m_S \gtrsim100$~MeV.

To calculate the branching ratio of the process, we utilize Heavy Quark Effective Field Theory~\cite{Neubert:1993mb}. First, we represent the matrix element of the decay by the process $q\bar{q} \to \gamma S$, where $V$ is assumed to be a bound state of $q\bar{q}$. Second, we assume that each quark $q,\bar{q}$ carries 1/2 of the $V$'s 4-momentum $p_{V}$. Third, we make the replacement 
\begin{equation}
\langle 0|\bar{q}\Gamma q | V\rangle \to \frac{f_{V}}{4}\text{Tr}[\slashed{\epsilon}\Gamma (m_{V}+\slashed{p}_{V})],
\end{equation}
where $f_{V}$ is the meson's decay constant that may be extracted from, e.g., the leptonic decays $V\to l \bar{l}$~\cite{Wilczek:1977zn}, while $\epsilon$ is the polarization vector.

Using these ingredients, for the tree-level matrix element, we have
\begin{multline}
    \mathcal{M}_{V\to \gamma S} = \sqrt{4\pi \alpha_{\text{EM}}}Q_{q}y_{q}\theta\cdot \frac{f_{V}}{4}\epsilon_{\sigma}(p_{V})\text{Tr}\left[ \gamma^{\sigma}\left(\frac{\gamma_{\mu}(\slashed{p}_{b}-\slashed{p}_{h}+m_{b})}{(p_{b}-p_{h})^{2}-m_{b}^{2}}+\frac{(\slashed{p}_{b}-\slashed{p}_{\gamma}+m_{b})\gamma_{\mu}}{(p_{b}-p_{\gamma})^{2}-m_{b}^{2}} \right)(m_{V}+\slashed{p}_{V})\right]\epsilon^{\mu}(p_{\gamma})
\end{multline}
Here, $Q_{q}$ is the quark's charge, while $y_{q} \equiv m_{q}/v$ is the Yukawa coupling. The squared matrix element averaged over $V$'s polarizations is
\begin{equation}
    \overline{|\mathcal{M}_{V\to \gamma S}|^{2}}=\frac{32\pi}{3} \alpha_{\text{EM}}\theta^{2}f_{V}^{2}Q_{q}^{2}y_{q}^{2}
\end{equation}
For the process $V\to \gamma SS$, we just need to replace $\theta \to\alpha/m_{h}^{2}$. 

The decay widths of these processes behave as (see also Ref.~\cite{Bauer:2022rwf})
\begin{align}
    \Gamma_{V\to \gamma S} &= \frac{2\alpha_{\text{EM}}\theta^{2}f_{V}^{2}Q_{q}^{2}y_{q}^{2}\left(1-x^{2}\right)}{3m_{V}},
\\
   \Gamma_{V\to \gamma SS} &= \frac{\alpha ^2 Q_b^2 y_b^2 \alpha_{\text{EM}} f_V^2 m_V \left(\sqrt{1-4 x^2} \left(2 x^2+1\right)+4 \left(x^4-x^2\right) \log
   \left(\frac{\sqrt{1-4 x^2}+1}{1-\sqrt{1-4 x^2}}\right)\right)}{48 \pi ^2 m_h^4},
\end{align}
where $x = m_{S}/m_{V}$. Next, let us express the decay constant $f_{V}$ in terms of the decay width $V\to e^{+}e^{-}$:
\begin{equation}
    \Gamma_{V\to e^{+}e^{-}} \approx \frac{4\pi\alpha_{\text{EM}}^{2}f_{V}^{2}Q_{q}^{2}}{3m_{V}}\Rightarrow f_{V}^{2} \approx \frac{3\Gamma_{V\to e^{+}e^{-}}m_{V}}{4\pi \alpha_{\text{EM}}^{2}Q_{q}^{2}}
\end{equation}
Finally, let us use the relation between the tri-linear coupling $\alpha$ and ${\rm Br}_{h\to SS}$ in the limit $m_{S}\ll m_{h}/2$ (c.f. Eq.~\eqref{eq:alpha-Br-h-SS}), $\alpha^{2} \approx 32\pi\text{Br}_{h\to SS}\Gamma_{h}m_{h}$, and replace $\Gamma_{V\to e^{+}e^{-}}=\Gamma_{V,\text{total}}\times {\rm Br}_{V\to e^{+}e^{-}}$. We get
\begin{align}
    \text{Br}_{V\to \gamma S} &\approx \text{Br}_{V\to e^{+}e^{-}}\frac{y_q^2 \theta^{2}}{2\pi \alpha_{\text{EM}}} \left(1-x^2\right), \\ \text{Br}_{V\to \gamma SS} &\approx \text{Br}_{h\to \text{SS}} \text{Br}_{V\to e^{+}e^{-}}\frac{\Gamma_h m_V y_q^2 \left(\sqrt{1-4 x^2} \left(2 x^2+1\right)+4 \left(x^4-x^2\right) \log \left(\frac{\sqrt{1-4
   x^2}+1}{1-\sqrt{1-4 x^2}}\right)\right)}{2 \pi^2 \alpha_{\text{EM}}
   m_h^3}
\end{align}

In the limit $m_{S}\ll m_{V}$, plugging all the numeric values for $\Upsilon(1S)$ and $J/\psi(1S)$ and using $\Gamma_{h} \approx 4\text{ MeV}$, we obtain
\begin{align}
    \text{Br}_{\Upsilon\to \gamma S} &\approx  1.9\cdot 10^{-4}\theta^{2}, \\ \text{Br}_{\Upsilon\to \gamma SS} &\approx  1.1\cdot 10^{-12}\frac{\text{Br}_{h\to SS}}{0.1}, \\ \text{Br}_{J/\psi\to \gamma S} &\approx  3.4\cdot 10^{-5}\theta^{2}, \\ \text{Br}_{J/\psi\to \gamma SS} &\approx  2.1\cdot 10^{-14}\frac{\text{Br}_{h\to SS}}{0.1}
\end{align}

The smallness of the branching ratios of 3-body processes makes it impossible to use them for searches. For example, we expect no more than $\simeq 10^{10}$ of $\Upsilon(1S)$ particles at Belle II. To obtain this number for SHiP, we launched a dedicated \texttt{PYTHIA8} simulation and found that $N_{\Upsilon}/N_{b\bar{b}} \simeq 10^{-2}$, which leads to $\sim 10^{12}$ of $\Upsilon$ particles produced during the full running time~\cite{Ovchynnikov:2023cry}.  The $J/\psi$ mesons present a similar suppression. 

The mono-production channels are much less suppressed, but their rates are controlled by the coupling $\theta$, which also enters the scalar lifetime. The latter is restricted by the requirement that scalar particles must propagate to the detector, i.e., $c\tau_{S}\gamma_{S} \gtrsim l_{\text{min}}$ where $l_{\text{min}}$ is the minimal displacement. The value $N_{V}\cdot \text{Br}_{V\to \gamma S}$ should be large enough in this region. 

For $m_{S} < m_{B}$, current constraints on $\theta$ impose $\text{Br}_{V\to \gamma S} \lesssim 10^{-11}$ rendering their contribution insignificant. However, for masses $m_{S}\gtrsim m_{B}$, one can have $\theta^2\sim10^{-4}$ and, thus, the channel $\Upsilon \to S+\gamma$ may contribute significantly. For SHiP, $l_{\text{min}} = 32\text{ m}$, and the minimal possible value of $\theta^{2}$ for $m_{S} \simeq 5\text{ GeV}$ to have comparable decay length is $\theta^{2}\simeq \text{few}\cdot 10^{-11}$. For Belle II, the situation is more optimistic, as the only cut is on the transverse displacement, $r_{\perp}>0.05\text{ cm}$. However, the whole chain 
\begin{equation}
\Upsilon(1S)\to S+\gamma \to \text{SM particles}+\gamma
\end{equation}
is more complicated to reconstruct due to the complicated hadronic decays of the scalar, and it is less clear against the background, as scalars with this mass would decay into a bunch of various states. We leave a detailed study of this question to future work.

\section{Experiments}
\label{app:experiments}

\subsection{SHiP}

\begin{figure}[h!]
    \centering
    \includegraphics[width=0.9\linewidth]{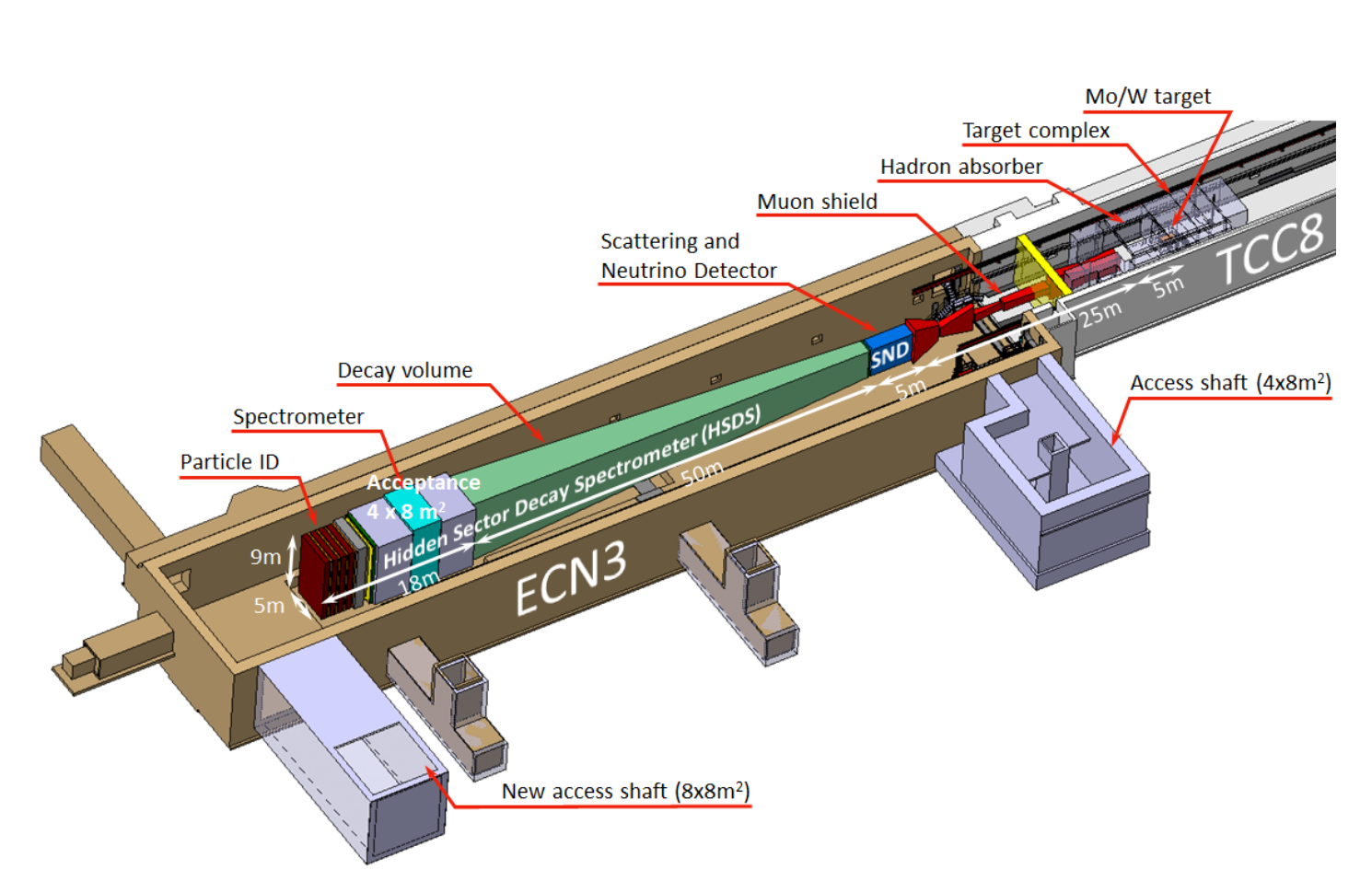}
    \caption{Setup of the SHiP experiment; the figure is taken from Ref.~\cite{Aberle:2839677}. In our studies, we consider smaller transversal dimensions of the decay volume and detector corresponding to the latest setup.}
    \label{fig:SHiP}
\end{figure}

The setup of the SHiP experiment we use mainly replicates the one discussed in~\cite{Aberle:2839677}, with the recent modifications of the transversal dimensions of the decay volume and detector, see Fig.~\ref{fig:SHiP}. We consider a 400 GeV proton beam hitting a beryllium target, with the total number of protons being $6\cdot 10^{20}$ during the 15-year operating time. The 50 m long decay volume is located 32 m downward of the target, being centered along the beam line, and has the form of an asymmetric pyramidal frustum with the dimensions
\begin{equation}
    \Delta x \times \Delta y = \begin{cases} 1\times 2.7\text{ m}^{2}, \quad \text{upstream.} \\ 4\times 6\text{ m}^{2}, \quad \text{downstream.} \end{cases} 
    \label{eq:transversal-dimensions-ship}
\end{equation}

The 11 m long detector system is located downwards the decay volume and contains a magnetized spectrometer with the integrated magnetic field $\int B\cdot dl = 0.65\text{ Tm}$, a timing system, and an ECAL. 

The amount of $B$ mesons produced during the full operating time of SHiP is~\cite{CERN-SHiP-NOTE-2015-009} $N_{B\bar{B}} \approx 1.5\cdot 10^{14}$; this number includes a cascade enhancement originating from secondary interactions of reaction products of the primary collision with the thick target. For the fragmentation fractions $f_{b\to B}$, we use~\cite{SHiP:2015vad,Alekhin:2015byh}
\begin{equation}
    f_{b\to B^{+}}=f_{b\to B^{0}}= 0.411, \quad f_{b\to B_{s}} = 0.11
\end{equation}

\subsubsection{Selection criteria and backgrounds}

Backgrounds at SHiP are made of the following contributions:
\begin{itemize}
    \item Muon combinatorial;
    \item Deep inelastic neutrino scattering;
    \item Deep inelastic muon scattering.
\end{itemize}
They occur either off the envelope of the decay volume or helium under atmospheric pressure filling it from inside.

Provided the pre-selection (with $\mathcal{O}(1)$ reconstruction efficiency) summarized below, as well as two background veto systems, the surrounding background tagger and the upstream background tagger, background becomes negligible: the upper bound on background yield is $0.3\text{ events}/15\text{ years}$~\cite{Aberle:2839677,SHiP:2025ows}.

For mono-decays, the pre-selection is summarized by the following criteria. The mono-decay event may be reconstructed provided that at least two decay products with total zero electric charge intersect the whole detector, have energies $E > 1\text{ GeV}$; the charged decay products must also have the transverse impact parameter no more than 2.5 m. We will utilize the same criteria for the di-decay events, which are also, obviously, background-free.

\subsection{Belle II}

\begin{figure}[h!]
    \centering
    \includegraphics[width=0.9\linewidth]{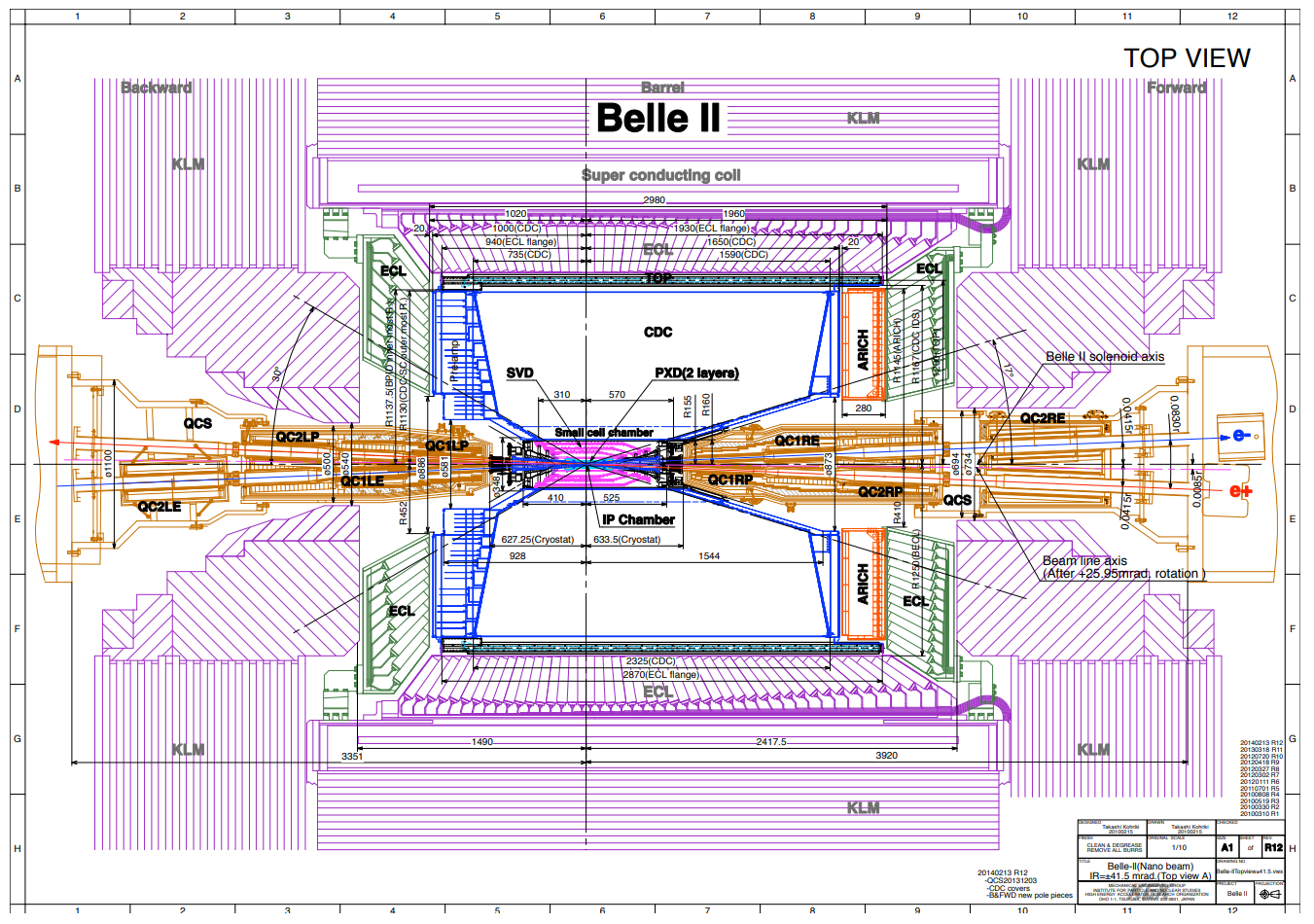}
    \caption{Setup of the Belle II experiment; the figure is taken from Ref.~\cite{Belle-II:2018jsg}.}
    \label{fig:BelleII}
\end{figure}

Belle II~\cite{Belle-II:2018jsg} is an electron-positron collider with a varying center-of-mass energy of the $e^{+}e^{-}$ pair, see Fig.~\ref{fig:BelleII}. The center-of-mass energy is chosen so that it closely matches the masses of heavy $b\bar{b}$-bound states, which subsequently decay into $B\bar{B}$ pairs. The main operating mode corresponds to $\sqrt{s_{e^{+}e^{-}}}$ closely matching the $\Upsilon(4S)$, decaying into $B^{+}B^{-}$ or $B^{0}\bar{B}^{0}$ pairs, with roughly equal probabilities of $51.4\%$ and $48.6\%$. 

The $e^{+}e^{-}$ collision is asymmetric -- the total momentum of the $e^{+}e^{-}$ pair has a small transversal component and a large longitudinal component:
\begin{equation}
p^{\mu}_{e^{+}e^{-}} = (11.006,0,0.125,3) \text{ GeV} \,.
\end{equation}
Because the electron and positron beam energies are asymmetric, the $e^{+}e^{-}$
center-of-mass frame is boosted in the laboratory. The detector geometry is correspondingly asymmetric: the collision point is shifted to the left relative to the center of the detector, leading to a larger acceptance of the events with the particles located in the forward direction. 

The full luminosity corresponds to $\mathcal{L} = 50\text{ ab}^{-1}$. Using information from Ref.~\cite{Belle-II:2023ueh} about the amount of $B\bar{B}$ pairs accumulated during the run with $\mathcal{L} =189\text{ fb}^{-1}$, which is $N_{B\bar{B}}^{189\text{ fb}^{-1}}=1.98\cdot 10^{8}$, we get $N_{B\bar{B}}^{50\text{ ab}^{-1}} \approx 5.24\cdot 10^{10}$. This is 4 orders of magnitude smaller than at SHiP.

\subsubsection{Selection criteria and backgrounds}

To study events with decaying FIPs at Belle II, we utilize the geometry and follow Ref.~\cite{Belle-II:2023ueh} as an example of a fresh collaboration analysis, performed for Higgs-like scalars. 

Unlike the case of SHiP, where we considered all the possible production mechanisms and scalar decay modes, for Belle II, we will restrict ourselves solely to the events in which the kinematics of the decaying $B$ meson can be fully reconstructed; this is needed to minimize the background. We will consider the decays
\begin{equation}
    B^{+}\to K^{+}+X/2X, \quad X \to Y^{+}Y^{-}, \quad Y = e,\mu,\pi,K
\end{equation}
and
\begin{equation}
    B^{0}\to K^{*0}+X/2X, \quad X \to Y^{+}Y^{-}, \quad Y = e,\mu,\pi,K,
\end{equation}
We drop the contributions from the $B$ decay into other resonances because they have a non-negligible decay width that would smear the kinematics of the reconstructed particles, introducing complexity in recovering the $B$ meson invariant mass in the event -- an essential feature required to suppress the background. Also, we drop the decays of scalars into the states with more than two particles.

The production mode $B_{s}\to XX$ may also be used at Belle~II. $B_{s}$ mesons may be produced via the decays of the $\Upsilon(5S)$ resonance appearing from the collisions of $e^{+}e^{-}$ particles at a larger $\sqrt{s_{e^{+}e^{-}}}$. As for the luminosity of the corresponding run, it would be reasonable to consider $\mathcal{L} = 5\text{ ab}^{-1}$, constituting the same fraction of the full luminosity as the dataset with $\Upsilon(5S)$ constituted at Belle. However, there is no kaon to tag in the $B_{s}$ decay, so quantification of background is required. We leave the investigation of to corresponding sensitivity for future studies.

Let us now specify the selection criteria:
\begin{itemize} 
    \item[--] The transverse displacement of the decay vertex $\mathbf{r}_{\text{dec}} = (x_{\text{dec}},y_{\text{dec}},z_{\text{dec}})$ must satisfy
    \begin{equation}
0.05\text{ cm}<\sqrt{x_{\text{dec}}^{2}+y_{\text{dec}}^{2}}<0.5\text{ m} \, .
    \end{equation}
    The lower bound corresponds to the minimal displacement used in Ref.~\cite{Belle-II:2023ueh}. The upper bound is caused by a sharp drop in the tracking efficiency at larger displacements.
    \item[--] The longitudinal displacement is mainly limited by the geometry of the CDC volume. We consider
    \begin{equation}
        -31\text{ cm}<z_{\text{dec}}<1\text{ m}, \quad 17^{\circ}<\theta_{\text{dec}} < 150^{\circ},
    \end{equation}
    where the angular cut follows from the geometry of the envelope of the beam pipes embedded in the CDC.
    \item[--] Charged kaons and pions from the prompt $B$ decays must have $p_{T}>0.15\text{ GeV}$. Their trajectories must intersect the detector system, which we assume to begin with ARICH detectors in Fig.~\ref{fig:BelleII}; i.e., the polar angles must range within $17^{\circ}<\theta_{K}<150^{\circ}$.
    \item[--] The trajectories of the displaced decay products must also intersect the detector system. Our strategy is to consider the 3-momentum of the decay product $(p_{x},p_{y},p_{z})$ and calculate the projection of its trajectory at the detector exit:
    \begin{equation}
        \mathbf{r}_{\text{proj}} = \left(x_{\text{dec}} +(z_{\text{exit}}-z_{\text{dec}})\frac{p_{x}}{p_{z}}, y_{\text{dec}} +(z_{\text{exit}}-z_{\text{dec}})\frac{p_{y}}{p_{z}}, z_{\text{exit}}\right),
    \end{equation}
    where $z_{\text{exit}}$ is either $1.65\text{ m}$ if $p_{z}>0$ (so the particle flies in the forward direction) or $-1.02\text{ m}$ in the opposite case. To surely intersect the detector, the radius $\sqrt{x_{\text{proj}}^{2}+y_{\text{proj}}^{2}}$ must be  $>60\text{ cm}$, where the ARICH system begins. If it is smaller, the decay product would escape the detector through the beam pipe surroundings.
    \item[--] Finally, the $p_{T}$ of all the displaced decay products must be $p_{T}>0.25\text{ GeV}$.
\end{itemize}

The backgrounds for the mono-decay search assuming the integrated luminosity $189\text{ fb}^{-1}$ have been discussed in Ref.~\cite{Belle-II:2023ueh}. We assume that for this luminosity, the typical background rate is $N_{\text{bg}}^{189\text{ fb}^{-1}}\simeq 10$ for each scalar mass. To obtain the background rate for the full luminosity run, we use the trivial rescaling 
\begin{equation}
N_{\text{bg}}^{50\text{ ab}^{-1}} = \frac{50}{0.189}N_{\text{bg}}^{189\text{ fb}^{-1}}
\end{equation}
Following the discussion in the main text, we assume zero background for the di-decay events.

To study the sensitivity of Belle II, we have conducted two independent simulations. The first one uses the sampler from Sec.~\ref{app:event-sampler}, where we also accommodated the selection based on the prompt kaon for the $B$ decays. The second one is an adaptation of our previous studies for the electron-positron colliders BaBar and Belle II in~\cite{Garcia:2024uwf}. As a cross-check of the approach, we have approximately reproduced the bounds on the Higgs-like scalars from Ref.~\cite{Belle-II:2023ueh}. The proper comparison is complicated, though. First, Ref.~\cite{Belle-II:2023ueh} utilized a different description of the scalar phenomenology, including, in particular, decays $S\to c\bar{c}$ right above the naive $cc$ threshold $m_{S} = 2m_{c}$ to the scalar lifetime. In reality, the true threshold for the corresponding decay is $m_{S}>2m_{D}$, where $m_{D}\approx 1.87\text{ GeV}$ is the $D$ meson mass. Second, the reported bound is wiggly, introducing an uncertainty within a factor of 2 in terms of the mixing angle $\theta$. 

\subsection{Downstream@LHCb}

\begin{figure}[h!]
    \centering
    \includegraphics[width=0.5\linewidth]{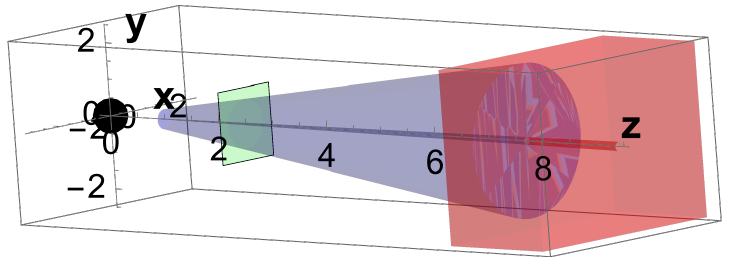}
    \caption{Geometry of the Downstream@LHCb setup, with all units in meters. The blue domain shows the pseudorapidity coverage of LHCb, $2<\eta<5$, starting from the longitudinal displacement $z = 1$ (the end of the inner tracker). The green plane shows the position of the UT tracker. Finally, the red domain defines the SciFi tracker. The figure is generated by \texttt{SensCalc}~\cite{Ovchynnikov:2023cry}.}
    \label{fig:Downstream@LHCb}
\end{figure}

The Downstream algorithm at LHCb~\cite{Gorkavenko:2023nbk,Kholoimov:2025cqe} allows reconstruction of the decay events of FIPs and long-lived SM particles such as $K_{L}$ and $\Lambda$ using the so-called Downstream tracks. These tracks leave hits only in two tracker sub-systems -- UT, located at $z \approx 2.5\text{ m}$, and SciFi, extended from $z = 7.75\text{ m}$ to $z = 9.4\text{ m}$. The effective decay volume extends in the longitudinal coordinate $z$ from $z = 1\text{ m}$ (the end of VELO) to $z = 2.5\text{ m}$ (the position of UT), see Fig.~\ref{fig:Downstream@LHCb}. The algorithm matches hits in UT and SciFi at the first High-level trigger (called HLT1) taking into account the magnetic field and uses neural networks to reduce fake tracks coming from spurious detector hits, with the efficiency of reconstruction at the level of $\epsilon_{\text{track}}\approx 80\%$ for the tracks with the transverse momentum $p_{T}>0.5\text{ GeV}$ and momentum $p>5\text{ GeV}$. 

The full reconstruction chain of the algorithm utilizes tracks' reconstruction, measuring tracks' momenta using the magnetic field with the bending power of 4 Tm, and vertexing, with the efficiency of $\epsilon_{\text{vtx}}\approx 90\%$ for the vertex made of two tracks, and suppressing background by using a neural network. The invariant mass may be subsequently reconstructed using the pion particle hypothesis. 

The algorithm was added to HLT1 in October 2024, and its performance has been validated by reconstructing the vertices of $K_{L}$ and $\Lambda$ from real data.

Ref.~\cite{Gorkavenko:2023nbk} studied the potential of Downstream@LHCb to explore various new physics scenarios. The setup has an excellent opportunity to explore the FIP parameter space in the near future. 

\subsubsection{Selection criteria and backgrounds}

We follow the selection criteria utilized in Refs.~\cite{Gorkavenko:2023nbk,Kholoimov:2025cqe}. First of all, we require the decaying particle to have the coordinates $1\text{ m}<z<2.5\text{ m}$ and $2<\eta<5$. We assume that only charged particles are reconstructible by the algorithm. For the track to be reconstructed, we require the particle's trajectory (affected by the magnetic field) to be within SciFi but not intersect the hole in the center, accounting for the beam pipe. The particle must have the transverse momentum $p_{T}>0.5\text{ GeV}$ and momentum $p > 5\text{ GeV}$. 

The background is expected to be under control for the fully reconstructible events with only two tracks. Various backgrounds have been studied in Ref.~\cite{LHCB-FIGURE-2025-002}, which combines simulations and real data acquired in October 2024, and found that the various backgrounds quickly drop as the invariant mass of the two reconstructed tracks increases, reaching a negligible level at $m_{\text{inv}}\geq 2\text{ GeV}$. Three types of background sources are considered:   
\begin{itemize}
\item \textbf{Hadronic Resonances}. Light hadronic resonances created from particle interactions with the beam pipe and detector materials can be suppressed by vetoing specific regions of the detector. Heavy resonances decay promptly inside the VELO detector, not reaching the Downstream region.
\item \textbf{Strange candidates}. $K_{S}$ and $\Lambda$ are vetoed in the mass region below 1.2 GeV.
\item \textbf{Combinatorial Background}. It comes from random combinations of tracks that fake a vertex, mainly low-momentum tracks. It is studied with simulated and real data events, using tracks with the same sign. This background decreases as the particle mass increases and can be removed using an NN trained against these events that have different characteristics from BSM signals. For very high masses, no contributions are expected.
\end{itemize}
The case of partially reconstructed events with more than two tracks (the typical decay mode of heavy Higgs-like scalars, see Fig.~\ref{fig:scalar-lifetime}) is currently under study. The di-decay signature is expected to be much less affected by this background since pairing two random fake FIPs does not provide the same origin vertex. Overall, we assume that the di-decay searches at Downstream@LHCb will be background-free. 

Therefore, to summarize, we conservatively require the events with at least two charged tracks per vertex passing the above-mentioned criteria, with the minimum Monte-Carlo-truth invariant mass $m_{\text{inv}}>2\text{ GeV}$. The overall event reconstruction efficiency is 
\begin{equation}
\epsilon_{\text{overall}} \approx \epsilon_{\text{track}}^{n}\cdot \epsilon_{\text{vtx}},
\end{equation}
where $n$ is the minimum number of tracks from the single particle decay with $m_{\text{inv}}>2\text{ GeV}$.

\subsection{Interpretation of the CMS constraints on di-decays}
\label{app:CMS-constraints}

We will utilize the constraint on di-decay events with 4 muons from Ref.~\cite{CMS:2024jyb}. Of most interest is the dataset from 2018, which excluded the parameter space of particles with the prompt lifetimes $c\tau_{X} < 100\text{ mm}$. We will use the top right panel of Fig.~5 of that paper, showing the model-independent constraint on the quantity
\begin{equation}
    \mathcal{S} = \sigma_{pp\to 2X}\cdot \text{Br}_{X\to 2\mu}^{2}\cdot \alpha_{\text{gen}}
\end{equation}
Here, $\sigma_{pp\to 2X} = \sigma_{pp\to h}\times \text{Br}_{h\to 2X}$, with $\sigma_{pp\to h}\approx 55\text{ pb}$~\cite{Cepeda:2019klc}, while $\alpha_{\text{gen}}(m_{X},c\tau_{X})$ is the model-independent selection efficiency, including the cuts on the position of the decay vertex as well as the kinematics and isolation of dimuons. 

Unfortunately, neither the mass nor the lifetime maps of $\alpha_{\text{gen}}$ have been provided in Ref.~\cite{CMS:2024jyb}, nor has it been applied to the Higgs-like scalar model. It is crucial to know the behavior of its efficiency -- we expect it to scale as $\alpha_{\text{gen}}\propto (c\tau_{X} \gamma_{X})^{-2}$ in the limit of large lifetimes, but the effects beyond this scaling may be also important. 

Therefore, we estimate its behavior using a simplified MC simulation, similar to the ones described in the previous sections. There, we decay the Higgs bosons produced in the $pp$ collisions at the LHC into a pair of the $X$ particles (taking the angle-energy distribution of the Higgs bosons from~\cite{Ovchynnikov:2023cry}), and apply the selection on the $X$'s decay vertices and dimuons from Table~1 of~\cite{CMS:2024jyb}.

\section{Background from accidental overlaps}
\label{app:pair-coincidences}

A potential concern is whether the di-decay signature could be faked by accidental overlap of unrelated events. There are two qualitatively different possibilities:
\begin{enumerate}
    \item an overlap between a genuine mono-decay and unrelated cosmic activity;
    \item an overlap of two independent mono-decays produced in beam-induced collisions.
\end{enumerate}
In this appendix we discuss both effects and show that they are negligible in the parameter space relevant for our study.

\subsection{Overlap of a mono-decay with cosmic activity}

We start with the possibility of an accidental coincidence between a genuine mono-decay and an unrelated cosmic-ray event. Such a fake di-decay candidate is strongly suppressed by a combination of timing, vetoing, and event topology.

For SHiP, the timing detector is expected to provide a resolution at the level of $\sigma_{\rm timing}\sim 100~\mathrm{ps}$~\cite{Aberle:2839677,SHiP:2025ows}. In addition, SHiP is equipped with dedicated veto systems surrounding and upstream of the decay volume (the surrounding background tagger and the upstream background tagger), specifically designed to reject entering charged particles and reduce non-beam backgrounds to a negligible level~\cite{Aberle:2839677,SHiP:2025ows}. The signal geometry also provides a strong handle: the LLPs are produced in the very forward direction, and the reconstructed momenta of their visible decay products point back to the beam line. Cosmic-ray tracks are not synchronized with the beam spill and are typically not aligned with this forward beam-induced topology. For the scalar case considered here, an accidental overlap would furthermore have to reconstruct two displaced vertices compatible with a common invariant-mass hypothesis, which provides an additional strong suppression.

The same qualitative conclusion applies to the other two experiments considered in this work. At LHCb, events are associated with the bunch-crossing structure of the LHC, with a 25~ns spacing between bunch crossings~\cite{LHCb:2014set}, and the reconstructed candidates are required to be compatible with a beam-induced event. Since Downstream@LHCb is a forward detector with a small solid-angle acceptance around the beam axis~\cite{Gorkavenko:2023nbk,Kholoimov:2025cqe}, cosmic-ray tracks are strongly disfavored by directionality. At Belle~II, the event timing is at the level of $\mathcal{O}(10~\mathrm{ns})$~\cite{Lai:2025gac}, the event environment is very clean, and trigger-level rejection of non-IP-like activity is provided, for example, by the CDC Level-1 $z$-vertex trigger~\cite{Bahr:2024dzg}. As a consequence, accidental overlap of a mono-decay with cosmic activity is not expected to fake the di-decay signal at any of the three experiments.

Since a fully realistic numerical estimate of cosmic-induced backgrounds would depend on detector-specific geometry, shielding, and veto performance, we do not attempt such a detector-level calculation here. Instead, we conclude that the combination of beam-time compatibility, veto systems, and the strongly constrained signal topology makes cosmic-induced fake di-decays highly implausible.

\subsection{Overlap of two independent mono-decays}

We now turn to accidental overlap of two independent mono-decays produced in beam-induced collisions. For concreteness, we consider SHiP as an explicit example and then comment on LHCb and Belle~II.

Let the mean number of reconstructed mono-decays per spill be
\begin{equation}
\mu \equiv N_{\rm mono,spill}
=
N_{B,\rm spill}
\,
\mathrm{Br}(B \to S + X)
\,
\epsilon_{\rm geom}
\,
P_{\rm decay},
\end{equation}
where $N_{B,\rm spill}$ is the number of produced $B$ mesons per spill. The total numbers of PoT and $B$ mesons during the 15-year running time of SHiP are $N_{\rm PoT}=6\times10^{20}$ and $N_{B}\sim 10^{14}$~\cite{SHiP:2015vad,SHiP:2025ows}. Each spill carries $N_{\rm PoT,spill}\sim 4\times10^{13}$, which corresponds to $N_{B,\rm spill}\sim 10^{7}$. The quantity $\epsilon_{\rm geom} \sim 0.2$ is a typical geometric acceptance at SHiP, including the decay volume and detector geometry~\cite{Ovchynnikov:2023cry}. Finally, $P_{\rm decay}$ is the probability for the scalar to decay within the fiducial region,
\begin{equation}
P_{\rm decay}
\approx
\exp\!\left[-\frac{L_{\min}}{\gamma c\tau_S}\right]
-
\exp\!\left[-\frac{L_{\max}}{\gamma c\tau_S}\right].
\end{equation}
For Higgs mixing, $\mathrm{Br}(B \to S+X) \sim 3\theta^2$~\cite{Boiarska:2019jym}, while $c\tau_S \propto 1/\theta^2$ is the scalar lifetime.

In the regime relevant to our exclusion contours, one has $\mu \ll 1$, so reconstructed mono-decays follow Poisson statistics. The probability of observing two independent mono-decays within the same spill is therefore
\begin{equation}
P(2) \simeq \frac{\mu^2}{2}.
\end{equation}

To fake a di-decay event, however, the two independent mono-decays must also be associated in time. This association is not controlled solely by the intrinsic timing capability of the detector. Indeed, for a genuine di-decay signal, the two scalars are produced simultaneously but decay at different positions inside the fiducial volume. Therefore, even for true signal events there is an irreducible time difference between the two reconstructed vertices,
\begin{equation}
(\Delta t)_{\rm Di\text{-}decay} = \frac{\Delta L}{\beta c} \simeq \frac{\Delta L}{c},
\end{equation}
where in the last step we used that the scalars are ultra-relativistic in the relevant SHiP parameter space. Since the maximal separation of the decay vertices is bounded by the size of the decay region, $\Delta L \lesssim 50~\mathrm{m}$, one finds the conservative upper estimate
\begin{equation}
(\Delta t)_{\rm Di\text{-}decay}
\lesssim
\frac{50~\mathrm{m}}{c}
\sim 10^{-7}~\mathrm{s}.
\end{equation}

At the same time, SHiP has a finite timing resolution, which we denote by $(\Delta t)_{\rm measurement} \sim n\,\sigma_{\rm timing}$, with $\sigma_{\rm timing}\sim 100~\mathrm{ps}$ and $n=\mathcal{O}(1)$~\cite{Aberle:2839677,SHiP:2025ows}. Therefore, the relevant effective association window for a di-decay candidate is
\begin{equation}
\Delta t_{\rm assoc}
\sim
\max\!\left[(\Delta t)_{\rm measurement},\,(\Delta t)_{\rm Di\text{-}decay}\right].
\end{equation}
For SHiP, the second contribution is the dominant one, so conservatively $\Delta t_{\rm assoc} \sim 10^{-7}~\mathrm{s}$.

Assuming that independent mono-decays are uniformly distributed over the spill duration $t_{\rm spill}$, the probability that two such decays fall within the same association window is approximately
\begin{equation}
P_{\rm assoc}
\simeq
\frac{\Delta t_{\rm assoc}}{t_{\rm spill}}.
\end{equation}
For SHiP, $t_{\rm spill}\sim 1~\mathrm{s}$, and hence $P_{\rm assoc} \sim 10^{-7}$.

The expected number of fake di-decay events over the full running time is therefore
\begin{equation}
N_{\rm fake}
=
\frac{\mu^2}{2}
\,
N_{\rm spills}
\,
\frac{\Delta t_{\rm assoc}}{t_{\rm spill}}.
\end{equation}
Since $\mu \ll 1$ throughout the parameter space relevant for our study, this suppression is sufficient to ensure $N_{\rm fake} \ll 1$ in the domain $\theta^{2}\lesssim 10^{-7}$ available for the future searches. Accidental overlap of independent mono-decays is therefore negligible.

\subsubsection{Slow versus fast scalars and timing considerations}

A related concern is whether the above estimate could be invalidated if one scalar is significantly slower than the other, thereby generating unusually large time-of-flight differences.

At SHiP, scalars are predominantly produced in decays of highly boosted $B$ mesons originating from 400~GeV proton collisions. In the forward geometry relevant for the decay volume, typical scalar boosts are $\gamma_S \sim \mathcal{O}(10\text{--}50)$ for $m_S \sim \mathcal{O}(1~{\rm GeV})$. Even near threshold in the $B$ rest frame, the lab-frame boost of the relativistic $B$ meson renders the scalar relativistic as well.

The time of flight over a baseline $L$ is $t = L/(\beta c)$, and for relativistic particles the delay relative to a particle with $\beta \simeq 1$ is
\begin{equation}
\Delta t \simeq \frac{L}{2c\gamma^2}.
\end{equation}
Taking $L \sim 50~{\rm m}$ gives
\begin{equation}
\Delta t \simeq \frac{83~{\rm ns}}{\gamma^2}.
\end{equation}
For representative boosts $\gamma = 10,\,20,\,50$, this yields time delays of order $0.8~{\rm ns}$, $0.2~{\rm ns}$, and $0.03~{\rm ns}$, respectively.

These values are completely negligible compared to the spill duration, and they remain well below the conservative signal-association scale $\Delta t_{\rm assoc}\sim 10^{-7}~\mathrm{s}$ used above. Therefore, varying scalar boosts do not invalidate the accidental-overlap estimate. To generate a time-of-flight difference comparable to the spill duration would require
\begin{equation}
\beta \sim \frac{L}{c\,t_{\rm spill}} \sim 10^{-7},
\end{equation}
corresponding to velocities of order tens of meters per second. Such kinematics are incompatible with GeV-scale particles produced in a high-energy beam dump.

At LHCb, the same argument is even stronger. The effective Downstream decay region extends only from $z=1~\mathrm{m}$ to $z=2.5~\mathrm{m}$~\cite{Gorkavenko:2023nbk,Kholoimov:2025cqe}, so the maximal separation of two genuine di-decay vertices is only $\Delta L\lesssim 1.5~\mathrm{m}$. Therefore the corresponding physical time splitting satisfies
\begin{equation}
(\Delta t)_{\rm Di\text{-}decay}^{\rm LHCb}
\lesssim
\frac{1.5~\mathrm{m}}{c}
\sim 5~\mathrm{ns},
\end{equation}
which is comfortably below the 25~ns bunch-crossing spacing of the LHC~\cite{LHCb:2014set}. In other words, genuine di-decay candidates naturally remain associated with the same bunch crossing. At the same time, the beam-induced overlap probability is much smaller than in the SHiP spill-based environment because the event intensity relevant for displaced LLP candidates is correspondingly lower. Accidental overlap of two unrelated mono-decays is therefore even less plausible at LHCb than at SHiP.

At Belle~II, the conclusion is stronger still. The event environment is essentially pileup-free in comparison with hadron-beam or hadron-collider experiments, while the event timing is at the level of $\mathcal{O}(10~\mathrm{ns})$~\cite{Lai:2025gac}. The probability of overlapping two independent LLP mono-decays is therefore negligible.

We therefore conclude that across the full parameter space considered, accidental overlaps -- whether between a mono-decay and cosmic activity or between two independent mono-decays -- cannot fake the di-decay signature studied in this work. Genuine di-decays, by contrast, arise from a common production time and exhibit correlated kinematics. Their two reconstructed vertices need not coincide within the intrinsic detector timing resolution; rather, they are naturally separated by $(\Delta t)_{\rm Di\text{-}decay} = \Delta L/(\beta c)$, with $\Delta L$ set by the relative positions of the two decay vertices. Timing information thus remains useful as an additional handle on the underlying production mechanism, while accidental overlaps of unrelated events stay negligible.

\section{Di-decay event sampler}
\label{app:event-sampler}

In this section, we describe the complex machinery we use to simulate di-decays of FIPs.

\subsection{\texttt{SensCalc}}

As a starting point, we use \texttt{SensCalc} -- a \texttt{Mathematica}-based code initially designed to calculate the event rate with decaying FIPs at various intensity frontier experiments~\cite{Ovchynnikov:2023cry}.\footnote{Available at \faGithub\href{https://github.com/maksymovchynnikov/SensCalc}{/maksymovchynnikov/SensCalc} or at \href{https://doi.org/10.5281/zenodo.7957784}{10.5281/zenodo.7957784}.} The code generates the tabulated angle-energy distribution of FIPs produced by various channels. Then, it computes the tabulated geometric acceptance for the FIP to decay inside the decay volume and the tabulated decay products acceptance -- a fraction of events satisfying geometric and kinematic cuts for the given experiment. For this, in particular, it samples the phase space of FIPs' decay products (if they do not include jets) or uses the phase space precomputed in \texttt{PYTHIA8}~\cite{Bierlich:2022pfr} (if jets are present). Finally, using the computed quantities, it integrates over all possible combinations of FIPs' and decay products' kinematics to obtain the event rate for mono-decays. \texttt{SensCalc} has been tested against various simulation frameworks and lightweight Monte-Carlo codes. 

Recently~\cite{Kyselov:2024dmi}, it has been extended by adding \texttt{EventCalc} -- a Monte-Carlo sampler that uses the tabulated angle-energy distribution of the FIPs generated by \texttt{SensCalc}, as well as the setups of various experiments, and then generates the decay vertices of FIPs inside the decay volume. Its \texttt{python}-based analog\footnote{Available at \faGithub\href{https://github.com/maksymovchynnikov/EventCalc-SHiP}{/maksymovchynnikov/EventCalc-SHiP}.} integrates \texttt{PYTHIA8} to simulate showering and hadronization of decay products on-flight and can be used to generate the FIP decay events in a format suitable for the software framework of the SHiP experiment. In another work~\cite{Garcia:2024uwf}, it has been adapted to sample the events in a model of inelastic dark matter coupled to a dark photon mediator, again using the tabulated distribution of dark photons precomputed by \texttt{SensCalc}.

\subsection{Modification to handle di-decays case}

\texttt{SensCalc} and its available modifications do not include routines to handle di-decays. Namely, at the very first step we need the correlated distribution of a pair of FIPs $X$ produced in the mother processes $Y \to Y' +2X$, whereas \texttt{SensCalc} uses the averaged distribution for one particle (saving the information about only one FIP in the case of di-production). However, we may use all of its machinery, including the database of various experiments, tabulated distributions of various mother particles, the phenomenology of different FIPs, and routines that sample the FIP production and decay chains, and modify them to incorporate di-decays.

Our modification incorporates the following logic (here not limited by the case of Higgs-like scalars):

\begin{enumerate}
    \item Fix the FIP and specify the description of its phenomenology and decay modes used to calculate the number of events (parametrize it by $\text{Br}_{\text{vis}}(m_{\text{FIP}})<1$). Then, only consider the production modes where the given FIP is produced in pairs. 
    \item Select the experiment, its setup, and selection criteria for decay products (detectable particles, their geometric and kinematic properties required for successful reconstruction).
    \item For a given FIP mass $m_{\text{FIP}}$, simulate the kinematics of $N_{\text{sim}}$ pairs of FIPs produced by decays of mother particles, i.e., their correlated 4-momenta given by $p_{\text{FIP}},\theta_{\text{FIP}},\phi_{\text{FIP}}$. After simulating, only keep those events for which both the FIPs are within the polar acceptance of the decay volume (parametrize it by $\epsilon_{\text{polar}}$, which can be either 0 or 1).
    \item Having the FIPs' kinematics and their proper lifetime $c\tau_{\text{FIP}}$, generate the positions of their decay vertices within the longitudinal displacement from the production point given by the position of the decay volume, and calculate the corresponding decay probability
    \begin{equation}
    P_{\text{dec}} = \exp\left[ -\frac{z_{\text{min}}}{c\tau_{\rm FIP}\beta_{\rm FIP}\gamma_{\rm FIP}\cos(\theta_{\text{FIP}})}\right]-\exp\left[ -\frac{z_{\text{max}}}{c\tau_{\rm FIP}\beta_{\rm FIP}\gamma_{\rm FIP}\cos(\theta_{\text{FIP}})}\right]
    \end{equation}
    At this step, leave only the events for which both FIPs from the generated pairs are simultaneously within the azimuthal range of the decay volume (parametrize it by $\epsilon_{\text{az}}$, which can be either 0 or 1). 
    \item Simulate the phase space of decays of both FIPs into various final states selected at the first step, starting from FIP's rest frame and then boosting it to the lab frame. For decays into jets, use the phase space pre-generated with the help of \texttt{PYTHIA8}. 
    \item For each decay, calculate the acceptance for the decay products $\epsilon_{\text{dec}}$ (which can be either $0$ or $1$) separately for each FIP from the produced pair. The latter is obtained by imposing requirements on detectable particles, geometric criteria, and kinematics. Afterward, leave only those events for which decays of both FIPs have $\epsilon_{\text{dec}} = 1$. 
\end{enumerate}
 
Let us introduce the quantity  $\xi^{(i,j)} = \epsilon_{\text{polar}}^{i,j}\cdot \epsilon_{\text{az}}^{i,j}\cdot P_{\text{dec}}^{i,j}\cdot \epsilon_{\text{dec}}^{i,j}$, where $i$ denotes the number of sampled event, and $j \le 2$ counts the FIPs produced per decay (for mono-production modes, it is equal to $1$, and we drop it). Using it, we may express the number of events for various signatures:
 
\begin{enumerate} 
\item Di-decay events:
\begin{equation}
    N_{\text{events}}^{(2)} = \sum_{Y}N_{Y}\times \text{Br}_{Y\to 2X}\cdot \frac{1}{N_{\text{sim}}}\sum_{i = 1}^{N_{\text{sim}}} \xi^{(i,1)}\cdot \xi^{(i,2)}
\end{equation}
where $i,1/2$ means the 1st/2nd particle of the $i$th pair of the FIP particles. 

\item (Mono+di)-decay events:
\begin{equation}
     N_{\text{events}}^{(1)+(2)} = \sum_{Y}N_{Y} \bigg[\text{Br}_{Y\to 2X}\cdot \frac{1}{N_{\text{sim}}^{(2X)}}\sum_{i = 1}^{N_{\text{sim}}^{(2X)}} \left(\xi^{(i,1)}+\xi^{(i,2)}-\xi^{(i,1)}\xi^{(i,2)}\right) +\text{Br}_{Y\to X}\cdot \frac{1}{N_{\text{sim}}^{(X)}}\sum_{i = 1}^{N^{(X)}_{\text{sim}}} \xi^{(i)}\bigg]
\end{equation}
Here, the first summand stands for the contribution of the di-production modes, whereas the second one describes the mono-production. $N_{\text{sim}}^{(X)}$, $N_{\text{sim}}^{(2X)}$ are the numbers of sampled events from mono- and di-production modes, with the property $N_{\text{sim}}^{(X)}+N_{\text{sim}}^{(2X)} = N_{\text{sim}}$. They are determined based on the ratio $N_{\text{prod}}^{X}$ and $N_{\text{prod}}^{2X}$:
\begin{equation}
    N_{\text{sim}}^{(2X)} = \frac{\text{Br}_{Y\to 2X}}{\text{Br}_{Y\to 2X}+\text{Br}_{Y\to X}}\cdot N_{\text{sim}}, \quad N_{\text{sim}}^{(X)} = \frac{\text{Br}_{Y\to X}}{\text{Br}_{Y\to 2X}+\text{Br}_{Y\to X}}\cdot N_{\text{sim}}
\end{equation}
\item Mono-decays only:
\begin{multline}
     N_{\text{events}}^{(1)} = \sum_{Y}N_{Y} \left[\text{Br}_{Y\to 2X}\cdot \frac{1}{N_{\text{sim}}^{(2X)}}\sum_{i = 1}^{N_{\text{sim}}^{(2X)}} \left(\xi^{(i,1)}(1-\xi^{(i,2)})+\xi^{(i,2)}(1-\xi^{(i,1)})\right) + \text{Br}_{Y\to X}\cdot \frac{1}{N_{\text{sim}}^{(X)}}\sum_{i = 1}^{N^{(X)}_{\text{sim}}} \xi^{(i)}\right]
\end{multline}
\end{enumerate}
As a working example, we have implemented the di-decays of Higgs-like scalars at SHiP, Belle II, and the Downstream algorithm. On top of what is described above, we have also incorporated the theoretical uncertainty in decays of the scalars following Ref.~\cite{Blackstone:2024ouf}, to demonstrate its impact on the sensitivities and constraints.

The code will be available with the next major update of \texttt{SensCalc}. It can be provided before this upon request. 

\subsubsection{Cross-checks}

We have validated the code for the mentioned setups by using two independent calculations:

\begin{enumerate}
    \item For the combined mono+di-decays signature in the domain $\theta^{2}$ where the di-$S$ production modes dominate (recall Fig.~\ref{fig:results}, top panel), it accurately reproduces the prediction of \texttt{SensCalc}.
    \item We also used a rough estimate of the number of events based on the formula~\eqref{eq:Nevents-di-supplemental} we defined in the main text. We recapitulate it here for completeness: 
    \begin{equation}
            N_{\text{events}}^{(2)} = N_{\text{prod}}^{2S}\times\left(\epsilon_{S}\cdot P_{\text{dec}}^{S} \cdot \epsilon_{\text{dec}}\right)^{2}, 
            \label{eq:Nevents-di-supplemental}
    \end{equation}
    Here:
\begin{itemize}
\item[--] $N_{\text{prod}}^{2S}$ is the total number of $S$ pairs originated from the di-production~\eqref{eq:di-production-scalar}.
\item[--] $\epsilon_{S}$ is the fraction of scalars whose trajectories intersect the decay volume. 
\item[--] $P_{\text{dec}}^{S}$ is the scalar decay probability:
\begin{equation}
P_{\text{dec}}^{S} = \exp\left[-\frac{l_{\text{min}}}{c\tau_{S}\langle\gamma_{S}\rangle}\right]-\exp\left[-\frac{l_{\text{max}}}{c\tau_{S}\langle\gamma_{S}\rangle}\right],
\label{eq:Pdecay-supplemental}
\end{equation}
with $l_{\text{min}/\text{max}}$ being the minimal and maximal distance from the collision point covered by the decay volume, and $\langle \gamma_{S}\rangle$ the mean scalar's $\gamma$ factor. 
\item[--]$\epsilon_{\text{dec}}$ is the fraction of the $S$ decay events that can be reconstructed; it includes the geometric part (aka the fraction of events where the trajectories of the minimal required number of the decay products are within the detector) and the reconstruction part (the suppression due to the lack of detectors and reconstruction efficiency).
\end{itemize}
For example, for the value of $\epsilon_{\text{dec}}\cdot \epsilon_{S}$ at SHiP, we used the value $0.2$, and $\langle E_{S}\rangle \approx 80\text{ GeV}$, which are the typical values for the mono-decay events~\cite{Ovchynnikov:2023cry}.

The predictions by this simple estimate in terms of the sensitivity are in reasonable $\mathcal{O}(2)$ agreement with our full simulation. The discrepancies are caused by the correlation between the kinematics of the scalars' pairs that are unaccounted for in the simple formula. 
\end{enumerate}

\section{Other models with trilinear $hXX$ coupling}
\label{app:other-models}

Apart from the Higgs-like scalars, we may consider other FIPs having trilinear coupling to the Higgs bosons, such as dark photons, axion-like particles (ALPs), and Heavy Neutral Leptons (HNLs). Similarly to the Higgs-like scalars at Downstream@LHCb, they may be produced by the decay $h\to SS$. In this section, instead, we briefly comment on the possibility of less trivial production $B,\Upsilon\to SS$, which is accessible at the facilities where the Higgs bosons cannot be produced.

We start with reminding the effective Lagrangian describing the interaction of scalars, Eq.~\eqref{eq:lagr-eff}:
\begin{equation}
    \mathcal{L}_{S} \supset \theta m_{h}^{2}hs+ \frac{\alpha_{S}}{2}hS^{2} \, ,
\end{equation} (further details in \cref{app:scalar-pheno}).

For the case of the dark photons $V$, one relevant model displaying the trilinear coupling is the Hidden Abelian Higgs Model (HAHM)~\cite{Curtin:2013fra,ATLAS:2022izj}. It has the effective Lagrangian
\begin{equation}
    \mathcal{L} \supset -\frac{\epsilon}{2}Z_{\mu\nu}V^{\mu\nu} + \alpha_{V} h V_{\mu}V^{\mu}, 
\end{equation}
where $V_{\mu \nu} = \partial_{\mu}V_{\nu}-\partial_{\nu}V_{\mu}$ is the $V$ field strength, and $Z_{\mu\nu}$ is the $U(1)_{Y}$ field. The first term gives rise to the phenomenology of the ``standard'' dark photon~\cite{Ilten:2018crw,Kyselov:2024dmi}. The second term originates from the mixing of the SM Higgs boson with the beyond-the-Standard-Model Higgs boson $h'$ with the mass $m_{h'} > m_{h}/2$. Parametrically, $\alpha_{V} = \tilde{\alpha_{V}}\cdot m_{V}^{2}$, which naturally eliminates unphysical longitudinal divergence of the matrix elements of the processes involving the $h\to VV$ transitions. 

As for the ALP case, we may have two qualitatively different effective interactions: when ALPs $a$ couple to the Higgs boson derivatively~\cite{Bauer:2017ris},
\begin{equation}
\mathcal{L}_{a,1} \supset \alpha_{a,1} h (\partial_{\mu}a)^{2}+\mathcal{L}_{a\Psi} 
\end{equation}
and non-derivatively~\cite{Dolan:2014ska}:
\begin{equation}
\mathcal{L}_{a,2} \supset \alpha_{a,2} h a^{2}+\mathcal{L}_{a\Psi} 
\end{equation}
Here, $\mathcal{L}_{a\Psi} $ denotes a set of operators involving only one $a$ field. 

In the HNL case, we consider the phenomenological Lagrangian~\cite{Graesser:2007yj,Graesser:2007pc,Caputo:2017pit,Butterworth:2019iff,Barducci:2020icf}
\begin{equation}
    \mathcal{L} \supset  m_N   U_{N}^i \, \bar{\nu}_iN+\alpha_N h\bar{N}^cN + \textit{h.c.}\, ,
\end{equation} 
where $\nu_i$, $i=1,2,3$, are the SM neutrino fields.

Note that the couplings $\alpha_{X}$, $X = N,V,a,S$ have different dimensionalities.

The $hXX$ interactions allow producing FIPs in the decays $h\to XX$. In addition, similarly to the case of the Higgs-like scalar, the $hXX$ term induces the decays
\begin{equation}
    B_{s}\to XX, \quad B^{+/0} \to Y_{s/d}+XX, \quad \quad B_{s}\to \phi + XX
    \label{eq:di-production-fip}
\end{equation}
We need to know whether the production rate governed by the processes~\eqref{eq:di-production-fip} is competitive enough to be considered for the di-decay signature.

We start with the averaged amplitude squared of the processes~\eqref{eq:di-production-fip} has the form
\begin{equation}
    \overline{|\mathcal{M}_{B\to Y_{D} XX}|^2} = \left|\frac{\xi_{b\to D}}{2v m_{h}^{2}}\right|^{2}\cdot\overline{\left|\mathcal{M}_{B\to Y_{D}}\right|^{2}}\cdot \begin{cases}
        \alpha_{S}^{2}, \quad \text{scalars}, \\ \alpha_{V}^{2}\cdot |(\epsilon(p_{V_{1}})\cdot \epsilon(p_{V_{2}}))|^{2}, \quad \text{dark photons}, \\ \alpha_{a,1}^{2}\cdot (p_{a_{1}}\cdot p_{a_{2}})^{2}, \quad \text{derivative ALPs}, \\ \alpha^{2}_{a,2}, \quad \text{non-derivative ALPs}, \\ \alpha_{N}^{2}\text{Tr}[v(p_{N_{1}})\bar{v}(p_{N_{1}})v(p_{N_{2}})\bar{v}(p_{N_{2}})], \quad \text{HNLs} 
    \end{cases} 
    \label{eq:scaling-B-XX}
\end{equation} 
Here, $\epsilon$ is the polarization vector of the dark photon, $p_{X_{1/2}}$ are the 4-vectors of the outgoing FIPs, $\xi_{b\to D}$ is the flavor changing neutral current coupling converting a $b$ quark to another down quark $D$ (corresponding to the transition $B\to Y_{D}$, and $\mathcal{M}_{B\to Y_{D}} = \langle Y_{D}|\bar{b}(1+\gamma_{5})D|B\rangle$ is the matrix element of the transition $B\to Y_{D}$~\cite{Boiarska:2019jym}. 

The next step is to replace 
\begin{align}
(\epsilon(p_{V_{1}})\cdot \epsilon(p_{V_{2}}))^{2} &\to m_{12}^{2}/m_{V}^{4} \sim m_{B}^{4}/m_{V}^{4}, \quad (p_{a_{1}}\cdot p_{a_{2}} )^{2}\to m_{12}^{2} \sim m_{B}^{4}, \quad  \, \\ \text{Tr}[v(p_{N_{1}})\bar{v}(p_{N_{1}})v(p_{N_{2}})\bar{v}(p_{N_{2}})] &\to m_{12} \sim m_{B}^{2}
\label{eq:replacements}
\end{align}  
The final ingredient is to express the coupling $\alpha_{X}$ in terms of $\text{Br}_{h\to XX}$, similarly to how we handled the Higgs-like scalar case (recall Eq.~\eqref{eq:alpha-Br-h-SS}). In the limit $m_{h} \gg 2m_{X}$, utilizing the same relations as in Eq.~\eqref{eq:replacements} but for the process $h\to XX$ (so $m_{12} =m_{h}^{2}$), we have
\begin{equation}
    \text{Br}_{h\to XX} \sim \frac{1}{\Gamma_{h}}\times\begin{cases}\alpha_{S}^{2}/m_{h}, \quad \text{scalars}, \\ \alpha_{V}^{2}m_{h}^{3}/m_{V}^{4}, \quad \text{dark photons}, \\ \alpha_{a,1}^{2}m_{h}^{3}, \quad \text{derivative ALPs}, \\ \alpha_{a,2}^{2}/m_{h}, \quad \text{non-derivative ALPs},  \\ \alpha_{N}^{2}m_{h}, \quad \text{HNLs}, 
    \end{cases}
\end{equation}
with $\Gamma_{h} \approx \Gamma_{h,\text{SM}}+\Gamma_{h\to XX} \approx \Gamma_{h,\text{SM}}$ being the decay width of the Higgs boson. 

Expressing $\alpha_{X}$ in terms of $\text{Br}_{h\to XX}$ and substituting these expressions in~\eqref{eq:scaling-B-XX}, we obtain
\begin{equation}
    |\mathcal{M}_{B\to Y_{D} XX}|^{2} = \left|\frac{\xi_{b\to D}\mathcal{M}_{B\to Y_{D}}}{vm_{h}^{2}}\right|^{2}\times\Gamma_{h}m_{h}\text{Br}_{h\to XX}\times \begin{cases}
        1, \quad \text{scalars}, \\ m_{B}^{4}/m_{h}^{4}, \quad \text{dark photons}, \\ m_{B}^{4}/m_{h}^{4}, \quad \text{derivative ALPs}, \\ 1, \quad \text{non-derivative ALPs}, \\ m_{B}^{2}/m_{h}^{2}, \quad \text{HNLs}
    \end{cases} 
    \label{eq:scaling-B-XX-2}
\end{equation}
It follows that in terms of the fixed $\text{Br}_{h\to XX}$, the probability of producing dark photons and derivatively coupled ALPs by the processes~\eqref{eq:di-production-fip} is heavily suppressed by the factor $(m_{B}/m_{h})^{4}\simeq 10^{-6}$ compared to the case of the Higgs-like scalars and non-derivatively coupled ALPs. The di-HNL production probability is suppressed weaker, $(m_{B}/m_{h})^{3}\simeq 10^{-3}$. 

Using Fig.~\ref{fig:production-probabilities} for the scalars and using these suppression factors, we may briefly conclude if the channels~\eqref{eq:di-production-fip} are relevant. The non-derivative ALPs do not have any relative suppression compared to the Higgs-like scalars. Therefore, they can be efficiently probed by the di-decay signature similar to the Higgs-like scalars. 

On the other hand, the di-production of dark photons and derivatively coupled ALPs in $B$ decays is heavily suppressed: utilizing the number of proton collisions $N_{p}$ at LHC and at SHiP, we have $N_{p}\cdot P_{\text{prod}} \lesssim 1$ for $\text{Br}_{h\to SS}\lesssim 0.1$. Therefore, the $B$ decays are irrelevant. For the setups considered here, this leaves Higgs decays at the LHC as the relevant source of dark photon or derivatively coupled ALP pairs.

It would be interesting to study this question for the HNLs, for which the relative suppression is there, but it is not as significant as for the dark photons and derivative ALPs.

\subsection{The Heavy Neutral Lepton Lagrangian}

Let us briefly introduce the HNL Lagrangian with this coupling consistently with the SM symmetries at high energies (following the approach of ref.~\cite{Bondarenko:2018ptm}): 
\begin{equation}
    \mathcal{L} \supset   y_N^i \, \bar{L}_i\tilde{H}N+ \dfrac{1}{\Lambda_N} H^\dagger H\bar{N}^cN + \textit{h.c.}\, ,
\end{equation} 
where $L_i$, $i=1,2,3$, are the SM lepton doublets and $\tilde{H}=\varepsilon_{lm}H^\ast_{m}$ is conjugated SM Higgs doublet. After the electroweak symmetry breaking, it is effectively reduced to the following Lagrangian: 
\begin{equation}
    \mathcal{L} \supset  m_N U_{N}^i \, \bar{\nu}_iN+\alpha_N h\bar{N}^cN + \textit{h.c.}\, ,
\end{equation} where $U_i = (v/m_N)y_i/\sqrt{2}$ and $\alpha_N = v/\Lambda_N$.  The first term describes the HNL mixing angles, whereas the second one is the trilinear coupling.

\section{Insights about di-decay signature}
\label{app:insights}

In this section, using the Higgs-like scalars at SHiP and the Downstream algorithm at LHCb as an example, we discuss some insights from simulating di-decay events and the opportunities delivered by observing such events.

\subsection{Di-decay events at SHiP}
\label{app:event-sampler-SHiP}

\begin{figure}[h!]
    \centering
    \includegraphics[width=0.5\linewidth]{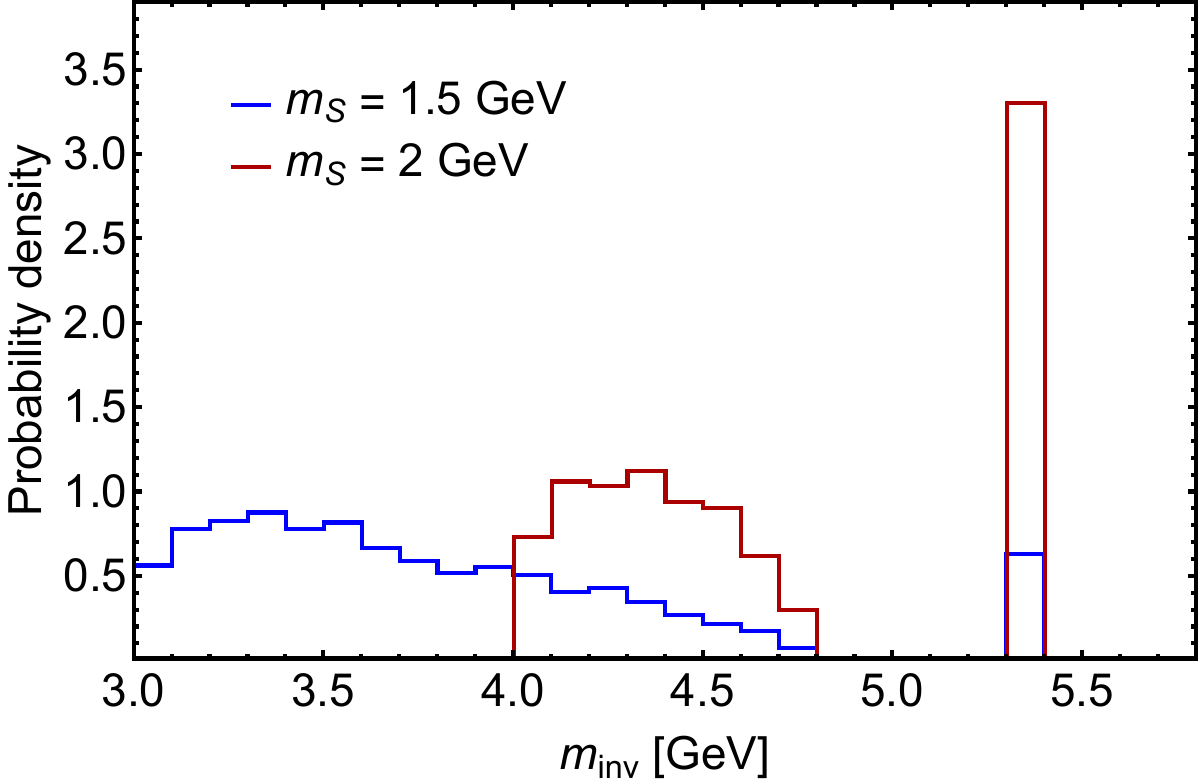}~\includegraphics[width=0.5\linewidth]{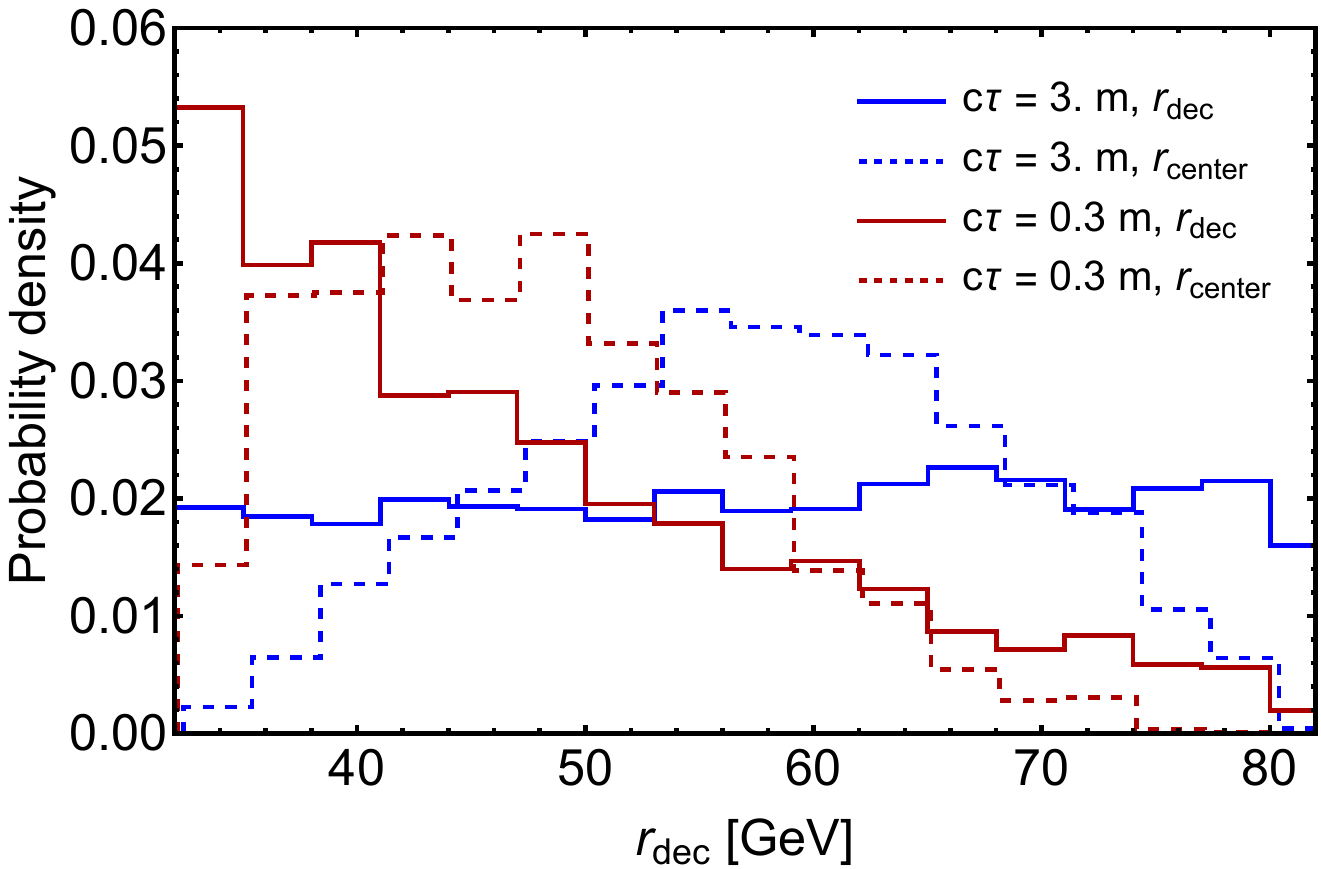}
    \caption{The Monte-Carlo-truth distributions of di-decay events of Higgs-like scalars at SHiP with the decay products passing the selection criteria. The distributions are obtained by weighting the reconstructed events with the decay probability of the scalars decaying inside the decay volume times the decay products acceptance. Left panel: combined invariant mass distribution of the pair of scalars. Two values of the scalar mass are considered: $m_{S} = 1 \text{ GeV}$ (the left panel) and $m_{S} = 1.7\text{ GeV}$ (the right panel), for the same value of the quantity $c\tau_{S}/m_{S}$, such that the scalars have the same decay length $c\tau_{S}\beta\gamma \simeq 40\text{ m}$. Right panel: the distributions in the decay vertex position for one of the scalars from the decaying pair ($r_{\text{dec}}$) and the ``center-of-decay'' $|\mathbf{r}_{\text{center}}|$, for the pairs of scalars of the same mass but different lifetimes.}
    \label{fig:minv}
\end{figure}

In the mass range of interest, $m_{S}< m_{B}/2$, decaying scalars have a relatively simple spectrum of decay modes -- main decays are two-body decays into a pair of charged particles. As a result, it is relatively simple to reconstruct the full kinematics of the di-decay events (the 4-momenta of the decaying scalars $p_{S_{1}}^{\mu}$ and $p_{S_{2}}^{\mu}$), and in particular the total invariant mass $m_{\text{inv}} = \sqrt{(p_{S_{1}}+p_{S_{2}})^{2}}$. The $m_{\text{inv}}$ distribution has two contributions: one from the 3-body decay $B\to X_{s/d}+2S$, and another one from the 2-body decay $B_{s}\to 2S$. I.e., it has a peak at $m_{\text{inv}} = m_{B_{s}}$, and a continuous distribution ending at $m_{\text{inv}} = m_{B}-m_{\pi}$. 

We provide an example of such a distribution for two different scalar masses in Fig.~\ref{fig:minv}. Given the excellent invariant mass resolution of SHiP -- around 20-40 MeV for GeV-scale particles, even 20-30 events may be enough for disentangling the 2-body and 3-body contributions and identifying the production modes of the scalar. By combining this information with the mono-decay events, we may identify all the interactions of the FIP leading to its production. 

Knowing the mother particle and the particular production modes, we may use the known energy spectrum of the production mechanism, convolve it with the exponential distribution governing the decay of each of the $X$\!s, and compare it with the observed distribution of di-decay vertices. Together with observing the mono-decay vertices and identifying the production modes, we may carefully extract the proper lifetime $c\tau_{\text{FIP}} = f(m_{\text{FIP}})/\theta^{2}$, and in particular disentangle the value of the coupling $\theta$ from the mass-dependent factor $f(m_{\text{FIP}})$. It is very important in light of significant theoretical uncertainties in the decay phenomenology of FIPs~\cite{Blackstone:2024ouf,Ovchynnikov:2025gpx}. Indeed, the di-decay events are much more sensitive to the proper lifetime than the mono-decays, as the di-decay event rate is proportional to $P_{\text{dec}}^{2}$.

As an illustration, Fig.~\ref{fig:minv} (right panel) shows the distribution of the ``center-of-decay'' position 
\begin{equation}
|\mathbf{r}_{\text{center}}| = \frac{1}{2}\sqrt{(x_{\text{dec},1}+x_{\text{dec},2})^{2}+(y_{\text{dec},1}+y_{\text{dec},2})^{2}+(z_{\text{dec},1}+z_{\text{dec},2})^{2}}
\label{eq:center-of-decay}
\end{equation}
of di-decay events versus the distribution of the decay vertex of mono-decay events; here, $\mathbf{r}_{\text{dec},i}$ denotes the decay vertex of the $i$th FIP from the decaying pair. The distributions are qualitatively different; for the parameter space where the mono-vertex distribution is flat, the di-decay vertex exhibits a non-trivial interplay between maximizing the decay probability and the decay products acceptance. 

\subsection{Di-decay events at Downstream@LHCb}

\begin{figure}[h!]
    \centering
    \includegraphics[width=0.33\linewidth]{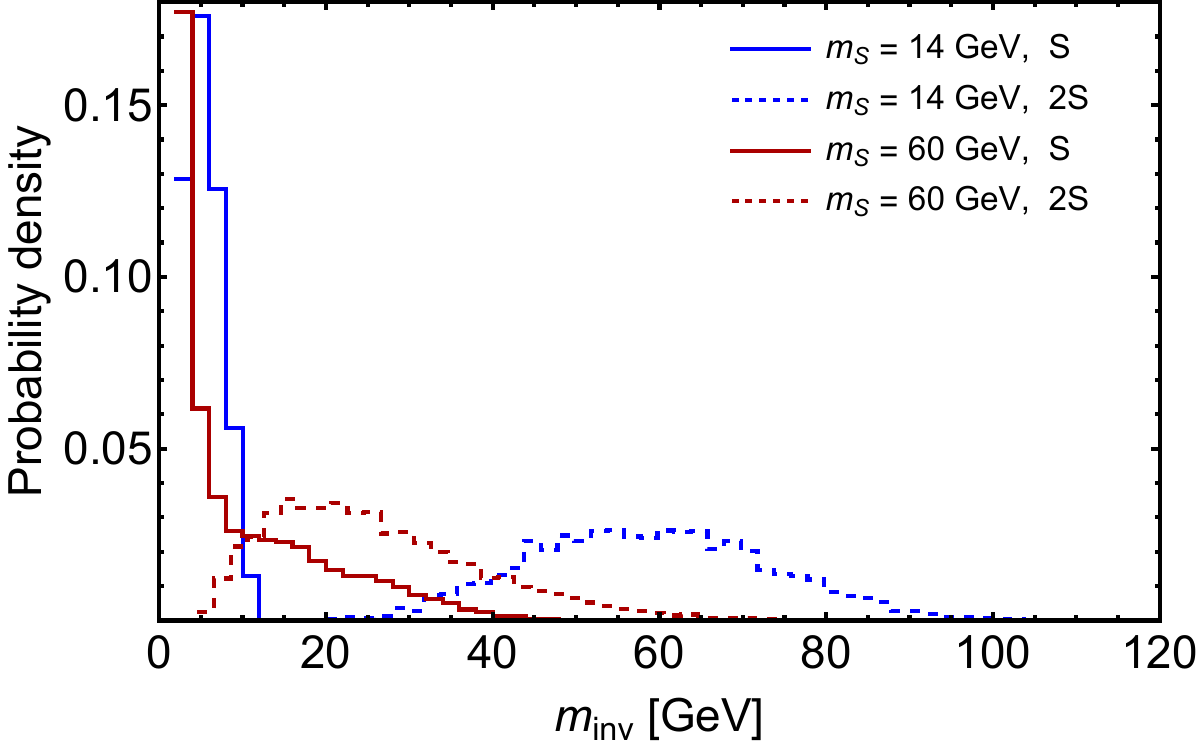}~\includegraphics[width=0.33\linewidth]{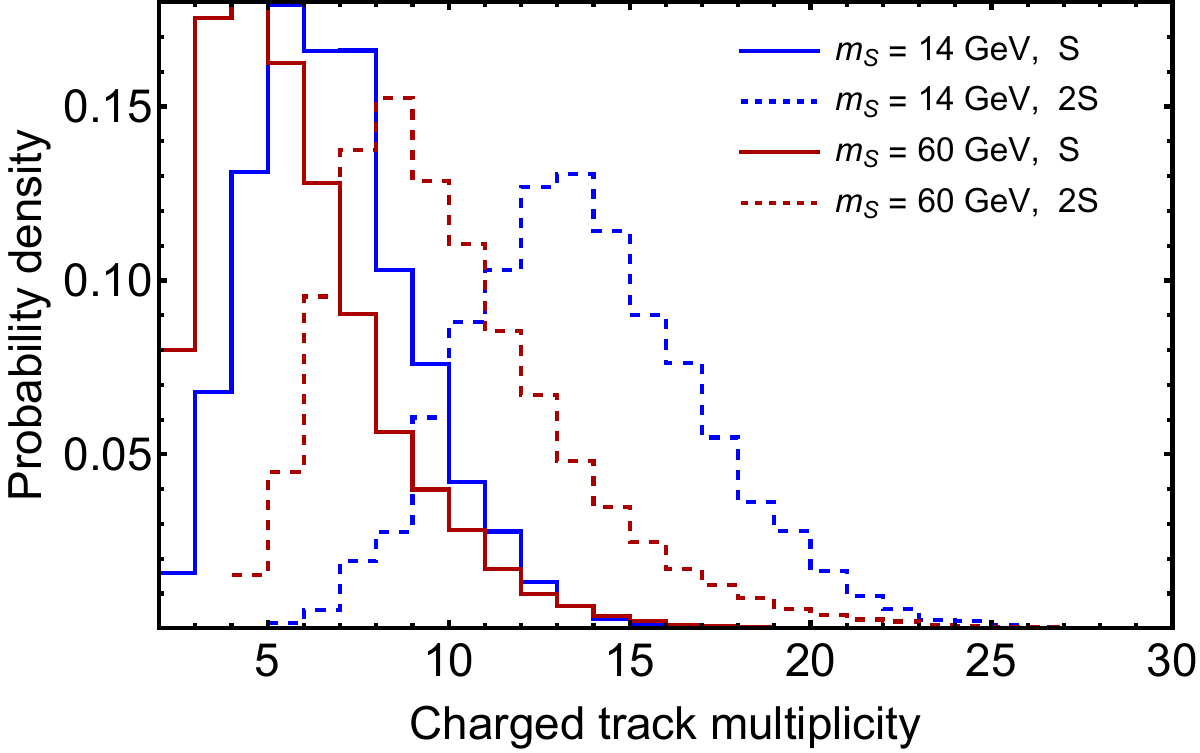}~\includegraphics[width=0.33\linewidth]{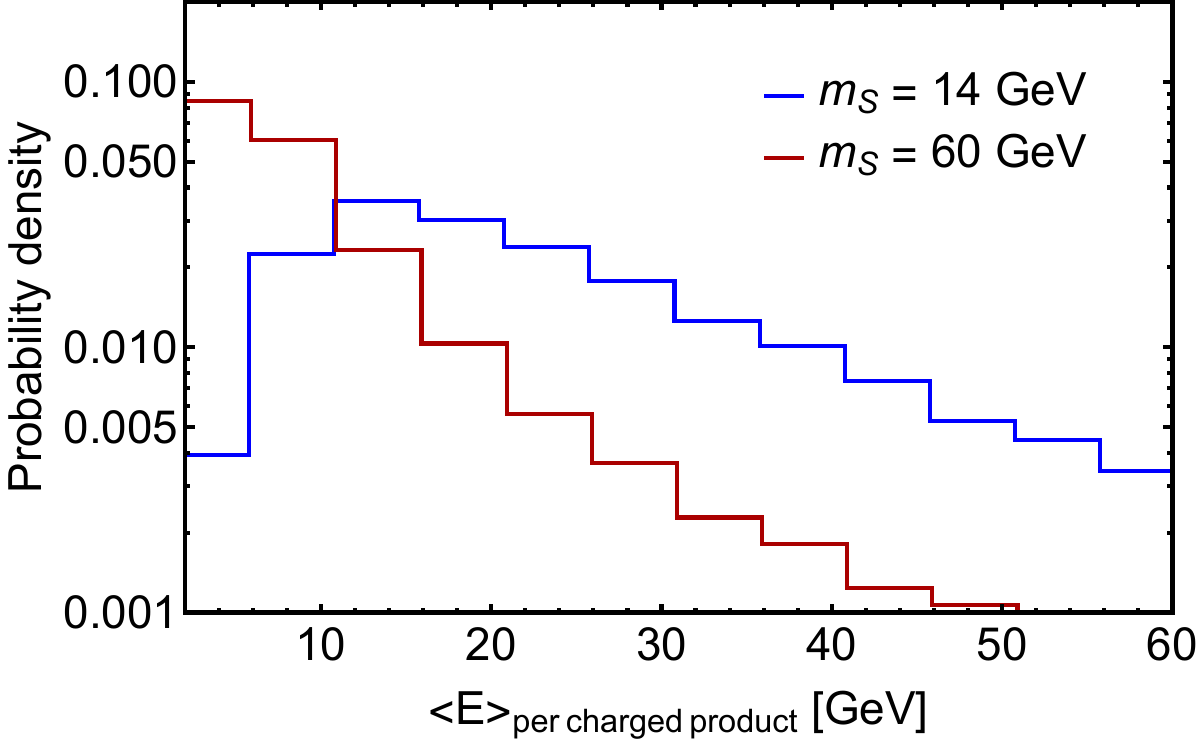}
\caption{The MC-truth invariant mass distribution (left), charged tracks multiplicity (middle), and the mean energy per charged decay product for the di-decay events with scalars at Downstream@LHCb. Two scalar masses are considered: $m_{S} =14\text{ GeV}$ and $60\text{ GeV}$, with the mixing angles for these two cases chosen such that the quantities $c\tau_{S}/m_{S}$ (and hence a typical decay length) are the same. Unlike Fig.~\ref{fig:results}, which is obtained by requiring at least two tracks with the invariant mass of $2$ GeV, the results in this figure are obtained by collecting all the charged decay products that pass the selection criteria (right). The lines denoted by ``S'' correspond to the quantities per single scalar from the decaying pair, whereas the ``$2S$'' lines denote the quantities where the information from both scalars is utilized.}
\label{fig:minv-multiplicity-Downstream}
\end{figure}

Let us now consider Downstream@LHCb. Fig.~\ref{fig:minv-multiplicity-Downstream} shows the MC-truth invariant mass and the multiplicity of all the tracks from di-decay events that pass the selection criteria, per one scalar and per two decaying scalars. Two reference scalar masses are considered: $m_{S} = 14$ and 60 GeV, with the same factor $c\tau_{S}/m_{S}$, such that they have similar typical decay lengths $c\tau_{S}\beta\gamma_{S} \simeq 1\text{ m}$ (i.e., the distribution of the decay vertices of each of the scalars in $z$ is close to homogeneous). The missing invariant mass is carried away by neutrinos and neutral pions, as well as by charged particles with low momenta. An interesting phenomenon is that increasing scalar mass is associated with decreasing multiplicity of reconstructed charged tracks and invariant mass of the decaying scalar pair. This is because most of the scalars decay into a $b\bar{b}$ pair. In subsequent showering and hadronization, they form a $B\bar{B}$ pair and a bunch of soft pions and kaons, which are not reconstructed by the algorithm. Both the multiplicity of these particles and the energy they carry away increase with the scalar mass. As the scalar energy remains the same in these processes, the mean energy per charged particle lowers, see Fig.~\ref{fig:minv-multiplicity-Downstream}. In addition, the $B\bar{B}$ pair becomes less aligned along the direction of the scalar's motion, which also influences the reconstructed invariant mass.

\begin{figure}
    \centering
    \includegraphics[width=0.5\linewidth]{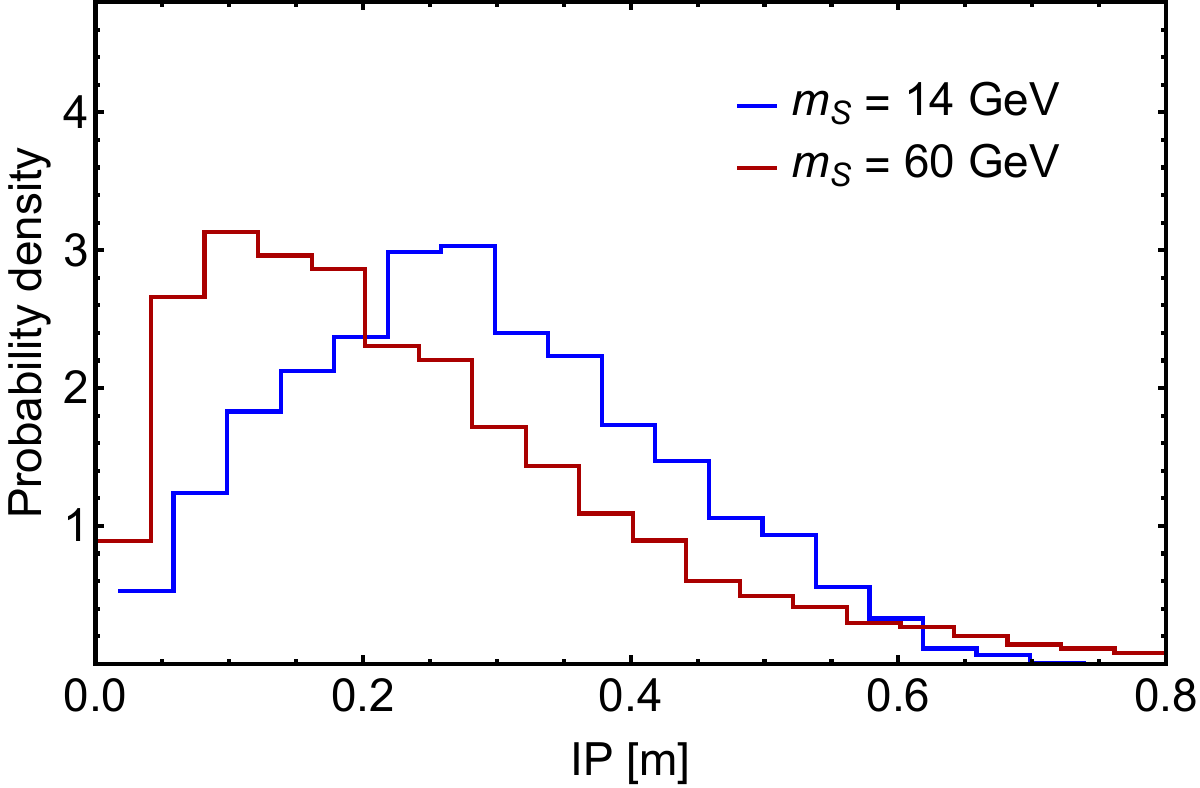}~\includegraphics[width=0.5\linewidth]{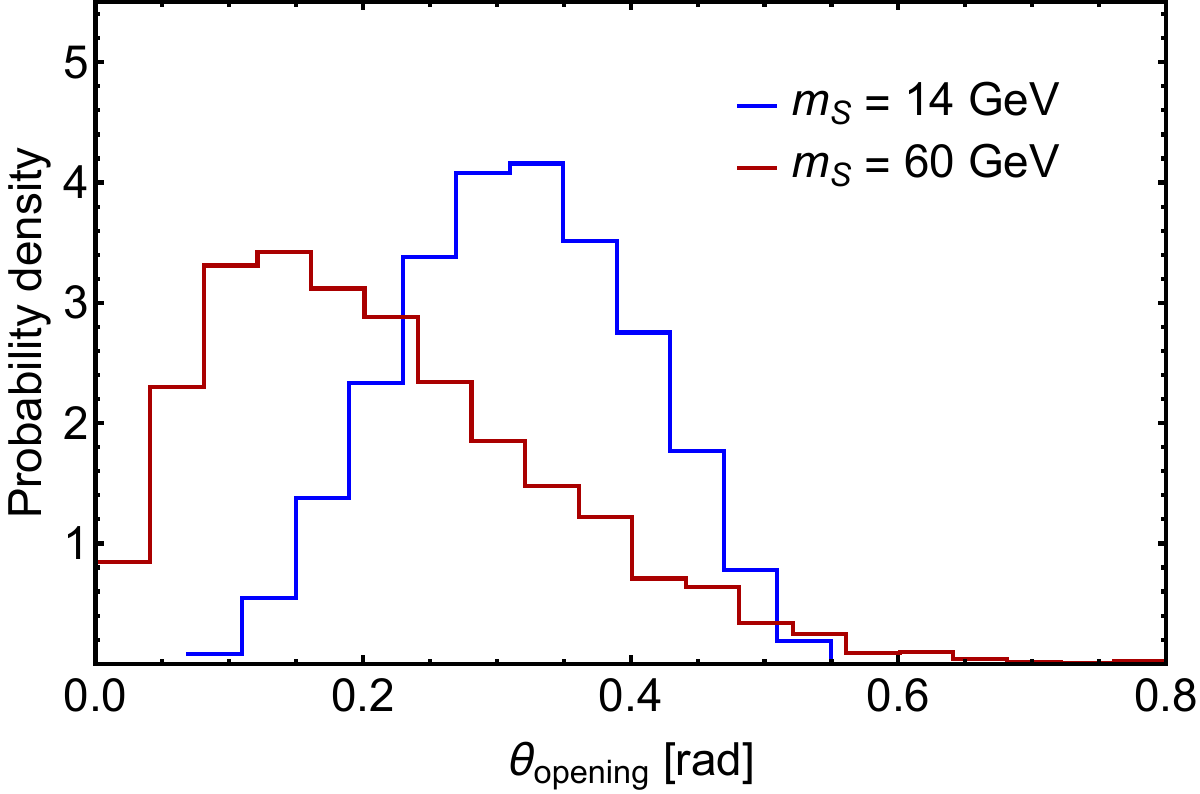}
    \caption{The distributions of the di-decay events in the transverse impact parameter with respect to the collision point for one scalar of the pair, and the opening angle between the two reconstructed 3-momenta. The same scalars as in Fig.~\ref{fig:minv-multiplicity-Downstream}.}
    \label{fig:IP-opening}
\end{figure}

Depending on the mass, the reconstructed events have a non-negligible transverse impact parameter with respect to the point of the proton-proton collision (IP), and a significantly varying opening angle between the two scalars, as calculated by the reconstructed momenta, see Fig.~\ref{fig:IP-opening}. They are larger for lower scalar masses. Indeed, the minimal possible opening angle behaves as $\sin(\theta_{\text{opening}}) \propto \sqrt{m_{h}^{2}-4m_{S}^{2}}$, i.e., it is much larger for the scalars with $m_{S}\ll m_{h}$. For large opening angles, any missing fraction of momentum would affect the IP more, which also explains a larger IP for lower mass.

These topological and kinematical distributions are very relevant to discriminating between signal and background events, both from material interaction and combinatorial background sources. In particular, the downstream track multiplicity for background events is only large in the domain of low invariant masses, coming from pion showers due to the interaction with detectors. That means that any signature of this type for the mass range above $\sim$\,2\,GeV is necessarily coming from new physics, being especially sensitive for the $2S$ case.

\end{document}